\documentclass[twocolumn]{aastex62}
\usepackage{amsmath,amssymb,amstext}
\usepackage{gensymb}
\usepackage{natbib}
\shorttitle{MOPPS II}
\shortauthors{May et al.}

\begin{document}
\title{MOPSS II: Extreme Optical Scattering Slope for the Inflated Super-Neptune HATS-8b}
\correspondingauthor{E. M. May}
\email{ermay@umich.edu}
\author{E. M. May}
\affil{University of Michigan \\
1085 S. University \\ 
311 West Hall \\ 
Ann Arbor, MI 48109-1107}
\author{T. Gardner}
\affil{University of Michigan \\
1085 S. University \\ 
311 West Hall \\ 
Ann Arbor, MI 48109-1107}
\author{E. Rauscher}
\affil{University of Michigan \\
1085 S. University \\ 
311 West Hall \\ 
Ann Arbor, MI 48109-1107}
\author{J. D. Monnier}
\affil{University of Michigan \\
1085 S. University \\ 
311 West Hall \\ 
Ann Arbor, MI 48109-1107}
\begin{abstract}
We present results for the inflated super-Neptune HATS-8b from MOPSS, The Michigan Optical Planetary Spectra Survey. This program is aimed at creating a database of optical planetary transmission spectra all observed, reduced, and analyzed with a uniform method for the benefit of enabling comparative exoplanet studies. HATS-8b orbits a G dwarf and is a low density super-Neptune, with a radius of 0.873 R$_{Jup}$, a mass of 0.138 M$_{Jup}$, and a density of 0.259 g/cm$^3$. Two transits of HATS-8b were observed in July and August of 2017 with the IMACS instrument on the Magellan Baade 6.5m telescope. We find an enhanced scattering slope that differs between our two nights. These slopes are stronger than one due only to Rayleigh scattering and cannot be fully explained by unocculted star spots. We explore the impact of condensates on the scattering slope and determine that MnS particulates smaller than 10$^{-2}\mu$m can explain up to 80\% of our measured slope if the planet is warmer than equilibrium, or 50\% of the slope at the equilibrium temperature of the planet for a low mean molecular weight atmosphere. The scattering slope that we observe is thus beyond even the most extreme case predicted by theory. We suggest further follow up on this target and host star to determine if the temporal variation of the slope is primarily due to stellar or planetary effects, and to better understand what these effects may be.
\end{abstract}
 
\section{Introduction} 
Our ability to probe the atmospheres of exoplanets is rapidly advancing, with transmission spectroscopy leading the way as a robust method to constrain the composition of a planet's atmosphere, for a well-characterized host star. With the number of detected exoplanets increasing rapidly in the past few years, we are no longer forced to study planets as single, unrelated, data points, but rather we are moving towards an era of comparative studies (see \cite{Sing2016} for a discussion of the variety of Hot Jupiter atmospheres, \cite{Fu2017} for an updated look at Hubble spectra, and \cite{Crossfield2017} for a look at Neptune-sized planets). In this new era for exoplanet science, it becomes necessary to have sets of uniformly observed and reduced planetary transmission spectra to better make comparisons between planets. In this work, we present our third data set in our catalog of such observations. Our work makes use of the IMACS instrument on the Magellan Baade telescope at Las Campanas Observatory in Chile.
\par Transmission Spectroscopy relies on detection of stellar light that filters through the planet's atmosphere as it transits the host star. The transit depth as a function of wavelength is then a combination of the light blocked by the optically thick part of the planet and the scattering/absorption effects on star light that passes through the atmosphere. When the stellar light is absorbed or scattered, the transit depth becomes larger, due to additional stellar light being removed from our line-of-sight. By precisely measuring the transit depth as a function of wavelength, we can probe the composition of the atmosphere based on absorption and scattering features; particularly allowing constraints on the mean molecular weight, molecular and atomic atmospheric constituents, and the presence of clouds or other scattering particles. The expected strength of a feature due to absorption and scattering scales with the planet's atmospheric scale height; a function of the planet's limb temperature divided by its surface gravity and atmospheric mean molecular weight. For these reasons, a typical `good' target for transmission spectroscopy is a planet with a high limb temperature, low atmospheric mean molecular weight, and/or a low gravity.
\par Observations of most exoplanet atmospheres with this method result in several seemingly distinct categories: `cloudy', exhibiting a flat and featureless transmission spectrum; `clear', showing evidence of atomic and/or molecular absorption with the full absorption profile present; or `hazy', showing evidence of scattering with few or diminished absorption features. At the optical wavelengths we probe in this work, the primary source of scattering is expected to be Rayleigh scattering, which has a wavelength dependence of $\lambda^{-4}$. This corresponds to a stronger effect at shorter, bluer, wavelengths, resulting in deeper transits as more light is scattered from our line-of-sight at the blue end of the spectrum. Meanwhile, the primary sources of absorption at optical wavelengths are the alkali metals Sodium and Potassium. The presence of their absorption profiles, and the exact shape, gives us information on the composition and properties of the planet's atmosphere.
\par Observations that only cover optical or infrared wavelengths are not always able to place unambiguous constraints on atmospheric properties alone. Therefore we are best able to characterize a planet when we combine data from multiple observations at a variety of wavelengths. The Hubble Space Telescope has been a leading force in infrared transit observations; from small, Neptune-like planets \citep[e.g.][]{Ehrenreich2014, Kreidberg2014}; to large, Hot Jupiters \citep[e.g.][]{Vidal2003,Sing2008}. More recently, \cite{Tsiaras2017} did a re-analysis of the Hubble observations for 30 exoplanets in order to provide a uniformly reduced and analyzed sample of planets. At optical wavelengths, numerous ground-based observatories are working on surveys; including the ACCESS group, also at the Magellan telescopes at Las Campanas, with results for GJ 1214 b \citep[]{Rackham2017} and WASP-19b \citep{Espinoza2018}; the Gran Telescopio Canarias exoplanet transit spectroscopy survey, currently with results for numerous planets \citep[see][]{GTC0,GTC1,GTC2,GTC3,GTC4,GTC5,Chen2017,GTC7,GTC8, GTC9}; GMOS at Gemini \citep[]{Huitson2017}; and FORS2 at the VLT \citep[]{Nikolov2016}. \cite{MOPSS1} (hereafter MOPSS1) is the first paper in our survey. Because telescope time is limited, it is important to make use of all resources available to us to study the atmospheres of new planets, as well as ensure reproducibility of results across telescopes, instruments, and reduction methods.
\par In this work, we present results for the low-density hot-Neptune HATS-8b from the Michigan Optical Planetary Spectra Survey (MOPSS) at the 6.5-meter Magellan Baade Telescope. MOPSS is designed with a goal of creating a catalog of uniformly observed and reduced transmission spectra to better enable comparative exoplanet studies. HATS-8b is the third target in the survey and was selected for its expected transmission signal, as well as the schedulability of transits at Magellan Baade.
\par In Section \ref{obs}, we discuss our observational set up and target. Section \ref{analysis} discusses our reduction pipeline in brief, as well as our noise model and generation of light curves. Section \ref{results} discusses our transmission spectra, as well as any impacts unocculted star spots may have on our results and the impact of particulates in the atmosphere. Finally, Section \ref{conclusions} presents our conclusions for this work.
%
\section{Observations} 
\label{obs}
\subsection{The Inamori-Magellan Areal Camera \& Spectrograph Instrument} 
IMACS (Inamori-Magellan Areal Camera \& Spectrograph) is a wide-field imager and optical multi-object spectrograph located on the 6.5-meter Magellan Baade Telescope at Las Campanas Observatory. We use the f/2 camera on IMACS and custom-made observing masks to simultaneously observe our target and a number of calibration stars. The IMACS f/2 CCD consists of 8 separate rectangular chips in a 4x2 arrangement. There is a gap of $\sim$57 pixels between the short edges of the chips, and $\sim$92 pixels between the long edges of the chips. 
\par We use the 300 lines/mm grism at a blaze angle of 17.5$^{\degree}$, resulting in theoretical wavelength coverage from 3900\AA{ }to 9000\AA, but practically, from $\sim$4600\AA{ } to $\sim$8000\AA{ } due to wavelength calibration; sources of noise (O$_2$ absorption in our atmosphere); and for the first transit, second order contamination beyond 8000\AA{ }. For the second transit, we use a blocking filter for light below 4550\AA{ }to eliminate this contamination below 9000\AA.
\begin{figure}
	\centering
    \epsscale{1.2}
    \plotone{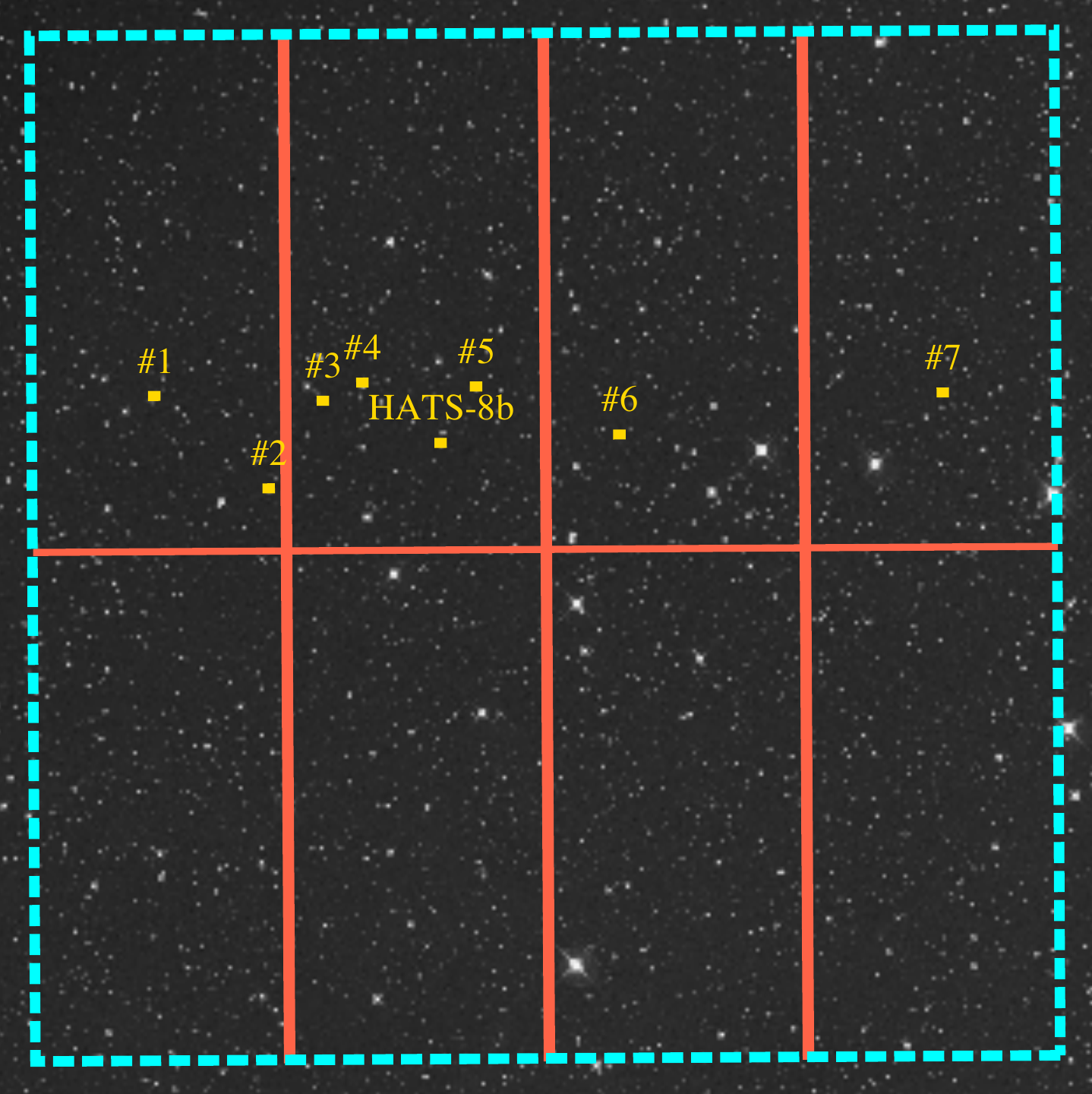}
    \caption{Field of View for HATS-8, with calibrator stars marked. The large yellow boxes represent the size of the slits cut on the instrument masks. The dispersion direction is in the vertical plane. Star \#3 serves as our `check star'.} \label{FOV}
\end{figure}
\par As mentioned above, we use a custom-made observing mask designed to maximize the number of calibrator stars, all aligned in a way to allow the full wavelength span to fall across the chips. The main target is placed on a central chip (see Figure \ref{FOV}). Calibrators were selected based on their magnitudes and spectral type if available, or their color in the common case that a spectral type was not determined. We select those stars that are most similar in color and are within $\sim$0.75 magnitudes of the main target when possible. Each calibrator and the main target is centered in a wide slit (15'') with large lengths (20'') so that we are not concerned with slit losses, and to improve background subtraction. With this setup, our resolution is limited by the seeing each night (see Sections \ref{Hats8_obs} for seeing data). Our observational efficiency was limited by the read-out time for the CCD; 82 seconds in fast mode and 1x1 binning, 29 seconds in fast mode and 2x2 binning).
\subsection{The planet HATS-8b} \label{Hats8_obs}
HATS-8b is an $\sim$1300K Hot Neptune orbiting a G-type star with a V-mag of 14.03. Discovered in 2015 \citep{Bayliss2015}, HATS-8b was immediately noted to be an ideal target for transmission spectroscopy. However, as of writing, there are no previous transmission spectra observations in the literature. 
\par HATS-8b was observed on the nights of July 23$^{rd}$ 2017 in 1x1 binning, with exposure times of 300 seconds and 72 exposures; and August 11$^{th}$ 2017 in 2x2 binning, with exposure times of 180 seconds and 92 exposures. This corresponds to an observational efficiency of 78\% and 86\%, respectively. Seeing was between $\sim$0.8'' and 1.5'' throughout the first night, with the seeing improving as the night went on. Our wide slits mean we were not concerned with slit losses, even at the 1.5'' seeing. We were not impacted by clouds or wind throughout the night. During the second night, seeing was around 0.6"-0.7" throughout the night with no cloud coverage. In addition to our main target, we observed an additional 7 calibrator stars which are listed in Table \ref{hats8cals}. We reserve 1 of the calibrators (star \#3 in Figure \ref{FOV} and Table \ref{hats8cals}) to serve as a 'check' that our calibrators have a constant flux throughout the night. This allows us to ensure that our pipeline is correctly accounting for airmass, seeing, and other instrumental effects throughout the night.
\begin{deluxetable}{cccccc}  
\tabletypesize{\small}
\tablecolumns{4}
\tablecaption{HATS-8b: Calibrator Stars \label{hats8cals}}
\tablewidth{0.4\textwidth}
\tablehead{
	\colhead{Identifier} 		& 
	\colhead{R.A.} 			& 
	\colhead{Dec.} 			& 
	\colhead{V mag.} 		&
	\colhead{R mag.} 		&
	\colhead{J mag.} 		}
\startdata
HATS-8   	&	19:39:46.08	&	-24:44:53.90	& 14.03	&  ---	&	13.10	\\
Cal \#1		&	19:40:14.20 &	-25:49:23.23	& 14.47 & 14.22	&	13.39	\\   
Cal \#2		&	19:39:56.82 &	-25:48:59.63    & 13.51 & 13.30	&	12.14	\\ 
Cal \#3		&	19:39:59.75	&	-25:46:15.68 	& 13.71 & 13.50	&	12.33	\\  
Cal \#4		&	19:39:57.51	&	-25:45:16.27	& 13.53 & 13.27	&	11.81	\\  
Cal \#5		&	19:39:47.79	&	-25:43:09.70	& 14.30 & 14.15	&	12.97	\\	
Cal \#6		&	19:39:31.70 &	-25:41:22.72    & 13.56 & 13.29	&	12.14	\\ 
Cal \#7		&	19:39:07.93	&	-25:34:29.86	& 14.46 & 14.21	&	13.41	
\enddata
\tablecomments{Simbad does not list an R magnitude for HATS-8. See Figure \ref{FOV} for a visual representation of the layout of the calibrator stars. Star \#3 serves as our `check star'.}
\end{deluxetable}
%
\section{Data Analysis} \label{analysis}
\subsection{Reduction Pipeline}
Our reduction pipeline was developed in Python and is described in detail in MOPSS1. It follows the traditional techniques for spectral reduction. We follow the process outlined in \cite{Nikolev2014} for cosmic-ray removal. We use SpectRes \citep{Carnall2017} for fast flux-conserving binning when down sampling the spectra to lower resolutions. 
\par Wavelength calibration is done using a second mask with small, 1'' square slits, at the locations of each object. We use a HeNeAr lamp and take several frames at the beginning of the night. By measuring shifts in the spectral direction as described in MOPSS1, we can shift the spectrum back to the wavelength frame of the first exposure, which should most exactly match the pixel locations of our arc frames. This paper includes improvements to our treatment of biases introduced due to atmospheric and instrumental effects, as described in the following sections.
\subsection{Removing Airmass Trend}
Because our red noise model is best applied to normalized data, we first correct for extinction due to airmass. This process requires fitting for extinction as a function of wavelength following the relationship
\begin{equation}
\label{airmass1}
    m_{observed}(\lambda,t)=m_{true}(\lambda,t)+k(\lambda)Z(t)
\end{equation}
where $m_{observed}$ is the observed magnitude at a given time and wavelength, $m_{true}$ is the true magnitude of the object, $k(\lambda)$ is the wavelength dependent extinction coefficient, and $Z(t)$ is the airmass as a function of time. We can use the observed magnitude and airmass from our first exposure and the definition of magnitudes to write Equation \ref{airmass1} as
\begin{equation}
\label{airmass2}
    F_{observed}(t)=F_{Z_0}10^{-k(\lambda)(Z(t)-Z_0)/2.5}
\end{equation}
where $F_{observed}(t)$ is the number of counts detected at time t, $F_{Z_0}$ is the number of counts at time=0, $k(\lambda)$ is the wavelength dependent extinction, $Z(t)$ is the airmass at time t, and $Z_0$ is the airmass at t=0. 
\par
\begin{figure}
	\centering
    \epsscale{1.15}
    \plotone{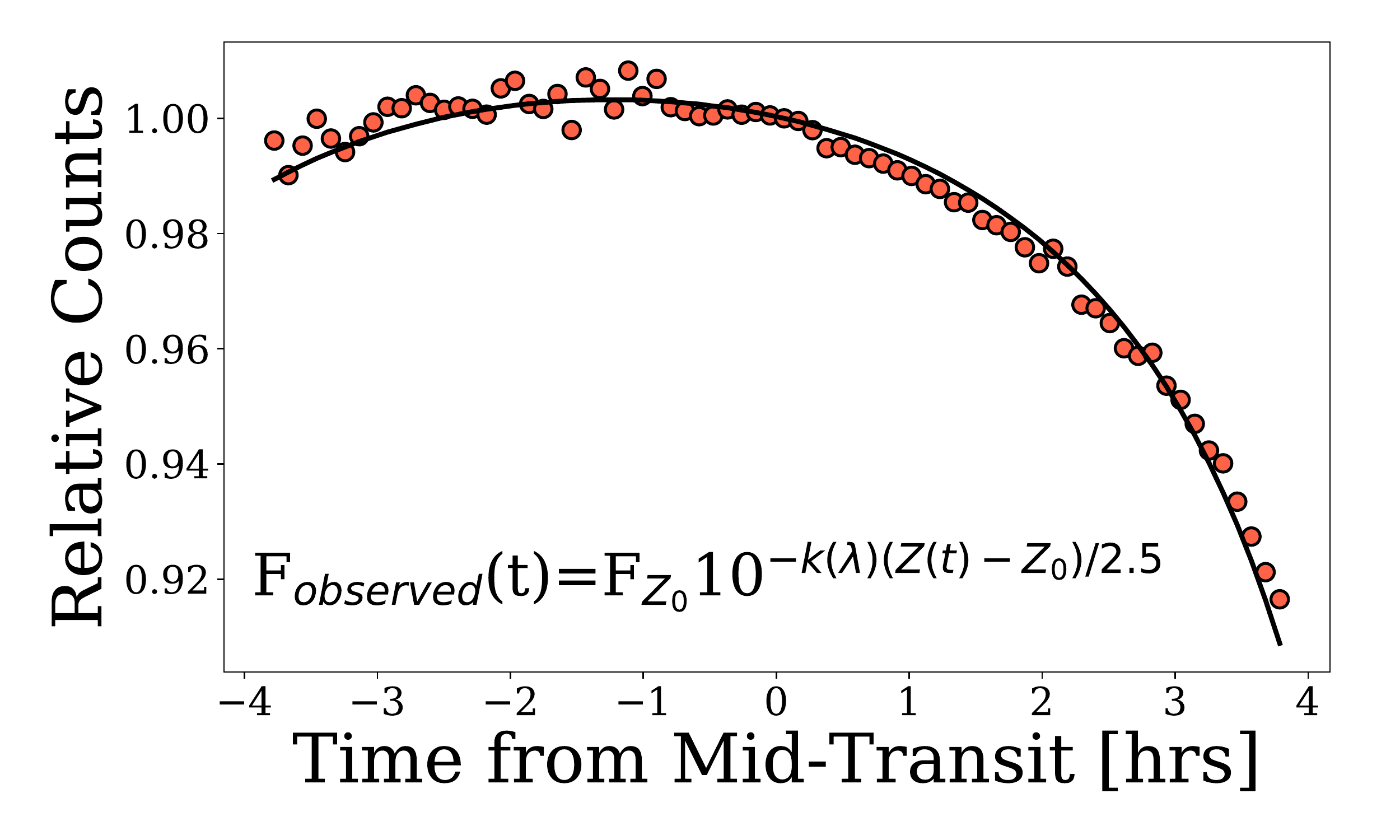}
    \caption{Here we plot raw, uncorrected, white-light data for our check star as well as an extinction curve derived from the mean extinction coefficient for all calibrator stars. This step removes the overall trend in the data, while keeping each object completely separate for our red-noise models.}\label{airmass_fit_fig}
\end{figure}

 We fit for the average $k(\lambda)$ across all objects except the target (the transit event will bias the fit), and then remove the trend described in Equation \ref{airmass2} from all objects. Figure \ref{airmass_fit_fig} shows an example of the extinction curve for the white light binning of our check star on night 1.
\par Typically, this trend is addressed simply through the division of the target star by the calibrator star(s). However, because we wish to model the correlated noise component in each star independently, we choose to remove the airmass trend in the manner described here. We find that treating each star independently in our noise model provides a greater precision in our final results. 
\subsection{Correlated Noise Model}
We expect the dominate source of correlated noise in our data to be caused by variations in seeing and the spectra shifting on the chips throughout the night. Though IMACS has very stable pointing, our night-long coverage of the object results in the objects shifting on the chips throughout the night at a measurable level of a few pixels. We extract these shifts during our data reduction process so that we can de-correlate the data against these shifts. The seeing is calculated by converting the spatial FWHM measured when flattening the 2D spectra to 1D, multiplied by the detector's plate scale of 0.2"/pixel.  
\par In addition to the spectral and spatial shifts and seeing, we de-correlate the data against the background counts as well. We use a linear combination of all of the above parameters fit to the out-of-transit times to predict in-transit data. Our model is described as
\begin{equation}
    f=\textbf{X}\beta
\label{noisemodel}
\end{equation}
where f is the data; X is a matrix containing the spatial and spectral positions relative to time=0, seeing, background counts, and a column of unity values to account for a constant offset; and $\beta$ is an array containing the relative importance of each term. By inverting Equation \ref{noisemodel},
\begin{equation}
   \beta=\left(\textbf{X}^{T}\textbf{X}\right)^{-1}f 
\end{equation}
one can solve for $\beta$ during the out-of-transit times, and use the results to calculate the predicted flux during transit under the current conditions (chip location, seeing, background).
\par
\begin{figure}
    \centering
    \epsscale{1.2}
    \plotone{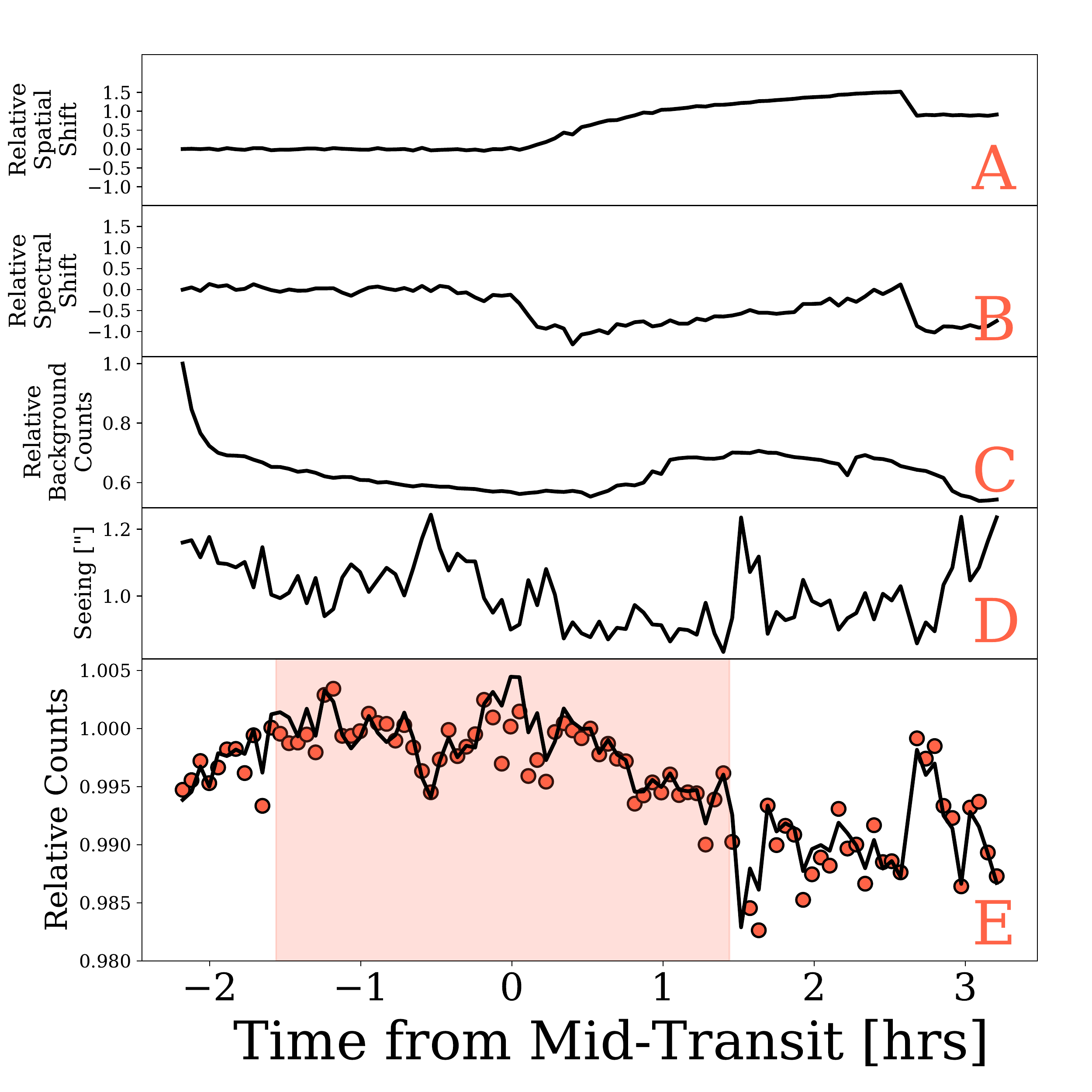}
    \caption{For our check star's white light binning, here we show the following: \textbf{A}: Pixel shift in the spatial direction as a function of time. \textbf{B}: Pixel shift in the spectral direction as a function of time. \textbf{C}: Background counts as a function of time. \textbf{D} Seeing as a function of time. \textbf{E} Airmass extinction corrected data with our red-noise model (see Equation \ref{noisemodel}) over plotted. The red box represents the in-transit times and are not used in the fit, but are predicted on.}
    \label{noisemodel_plot}
\end{figure}

 We perform this fit for each object and bin independently. Though relative pixel shifts are the same for all objects, due to differences in pixel response, a shift of 1 pixel in the spatial direction may mean an increase in counted photons for one object or wavelength bin, but a decrease for another object or wavelength bin. Figure \ref{noisemodel_plot} shows an example of our noise model as applied to the white light data for our check star on night 2. We are able to explain the majority of the trends and scatter with this method. Any trends left over we attribute to instrumental effects we are unaware of or do not include. Theses trends are small in comparison to what we fit for here, and are removed through dividing our target by a master calibrator star. 
\subsection{Light Curves}
To generate our light curves, we first create our `master' calibrator which consists of all reference stars except the one chosen as the `check' star. The master calibrator allows us to correct for instrumental effects not captured in our noise model. After applying the master calibrator, a small systematic generally remains in the light curve. Though this trend is smaller than in previous work due to our airmass and red noise corrections, we still find it necessary to include a baseline fit. For night 1, we fit this as a 3$^{rd}$ order polynomial in time, as detailed in MOPSS1. For night 2, we use a 2$^{nd}$ order polynomial in time due to there being relatively few data points before transit to constrain the fit. 
\subsubsection{White Light Curves}
The white light curves for HATS-8b are generated from 4200-8000\AA{} on night 1 and 4600-9000\AA{} on night 2. The difference is due to a blocking filter we implemented on night 2 as described in Section \ref{Hats8_obs}. The white light curves are used to fit for orbital parameters in the same manner as in MOPSS1. We use Batman \citep{Kreidberg2015} with a quadratic limb-darkening function, as well as emcee \citep{FMackey2013} to fit for center of transit, period/semi-major axis, inclination of orbit, white-light planet radius, and white-light limb darkening. All orbital parameters are fit with a Gaussian prior defined by the values and errors given in Table \ref{Hats8_params}. We find that our MCMC chains do not converge as easily if we have both the period and semi-major axis as free parameters, and so we maintain the relation between period and semi-major axis by only fitting for period while simultaneously calculating a corresponding semi-major axis based on the stellar mass and radius and their errors cited in Table \ref{Hats8_params}.
\begin{deluxetable}{cc} 
\tabletypesize{\small}
\tablecolumns{5}
\tablecaption{HATS-8b: Stellar and Orbital Parameters \label{Hats8_params}}
\tablewidth{0.8\textwidth}
\tablehead{
	\colhead{Stellar Parameter}		&
	\colhead{Value}			           	}
\startdata
Mass, [M$_{\odot}$]			&	1.056$\pm$0.037		\\
Radius, [R$_{\odot}$]		&	1.086$^{+0.149}_{-0.059}$  	\\
T$_{eff}$, [K]				&	5679$\pm$50		\\
{ }[Fe/H], [dex]				&	0.210$\pm$0.080		\\
$\xi$, [km/s]				&	2.00$\pm$0.50	\\
{\scriptsize{(Microturbulent Velocity)}} & { }				\\
$\log{g}$, [cm/s]			&	4.386$\pm$0.071		\\ \hline \hline
Planet Parameter			&	Value	\\ \hline
Mass, [M$_{Jup}$] 	& 0.138$\pm$0.019				\\
Radius, [R$_{Jup}$] 	& 0.873$^{+0.123}_{-0.075}$				\\
T$_{eq}$, [K]			& 1324$^{+79}_{-38}$				\\
Period, [days]		& 3.583893$\pm$0.000010			\\
Semi-major axis, [AU]	& 0.04667$\pm$0.00055		\\
Eccentricity			& $\textless$0.376							\\
Inclination, [deg]		& 87.8$^{+1.2}_{-1.8}$		\\ 				
\enddata
\tablecomments{All values have been adopted from \cite{Bayliss2015}. Our fits use an eccentricity of 0.0 due to the lack of strong constraint from previous work.}
\end{deluxetable}
\subsubsection{Binned Light Curves}
\begin{figure*}
	\epsscale{1.2}
    \plotone{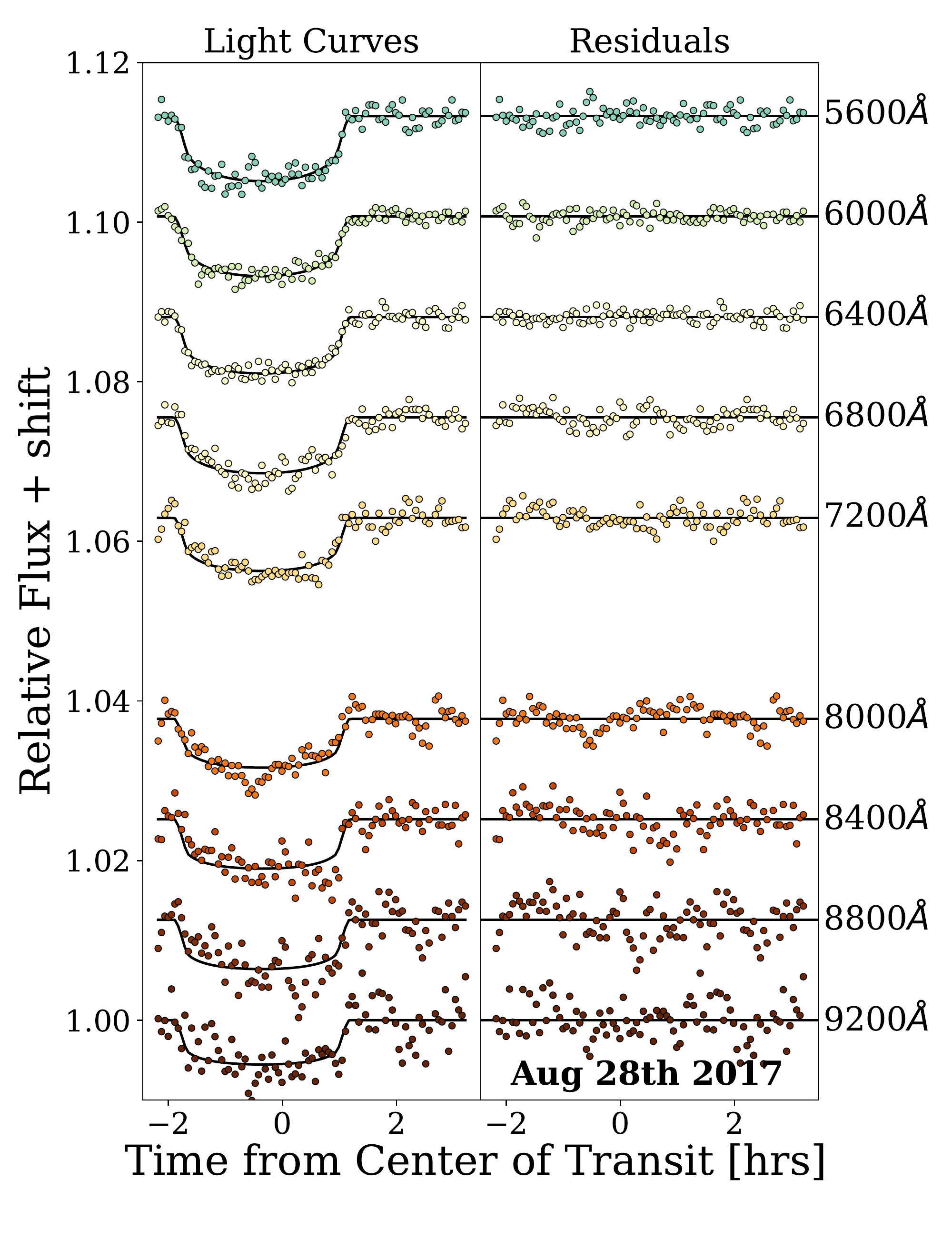}
    \caption{Light curve data and MCMC fits. \textbf{Top:} Results from the night of July 23rd 2017. \textbf{Bottom:} Results from the night of August 28th 2017. For both, the left panel shows the data and fits, and the right panel show the residuals (data-model).} \label{LCS}
\end{figure*}

We bin the data from each night into bins of width 400\AA, 200\AA, 100\AA, 50\AA, 40\AA, and 20\AA{ }in order to capture the overall shape of the transit at high precision, while also searching for absorption features that may be averaged over in wider binnings. For each resultant light curve, we use emcee to fit for $R_{p}/R_{S}$, and the two quadratic limb darkening parameters. We do not fit for orbital parameters as a function of wavelength as they should not have a wavelength dependence. We use the orbital parameters we retrieve from our white light emcee runs. Figure \ref{LCS} shows the 400\AA{ }binned light curves and their fits from our MCMC runs.
\begin{figure*}
	\centering
    \epsscale{1.2}
    \plotone{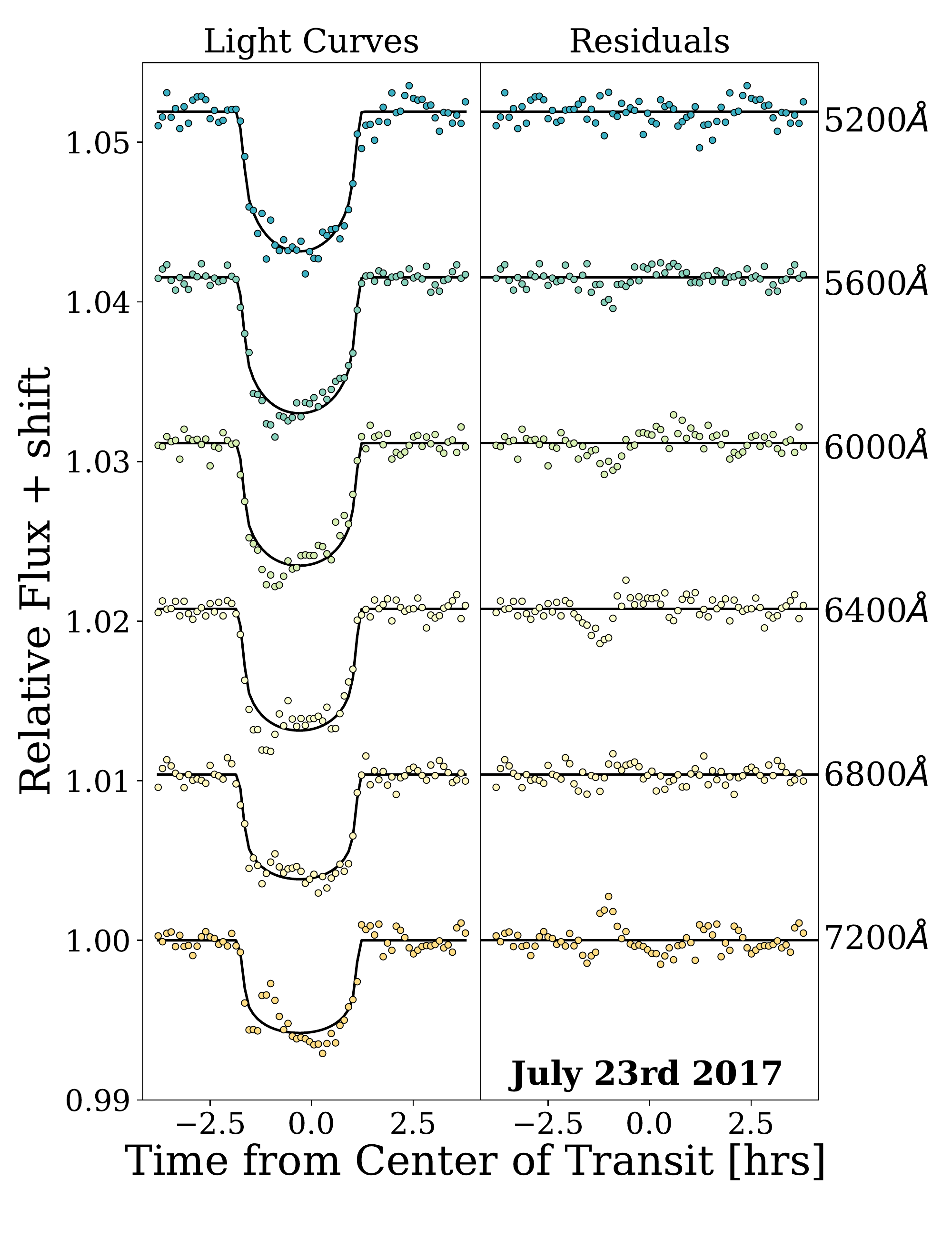}
\end{figure*}
\par
\begin{figure}
	\centering
    \epsscale{1.2}
    \plotone{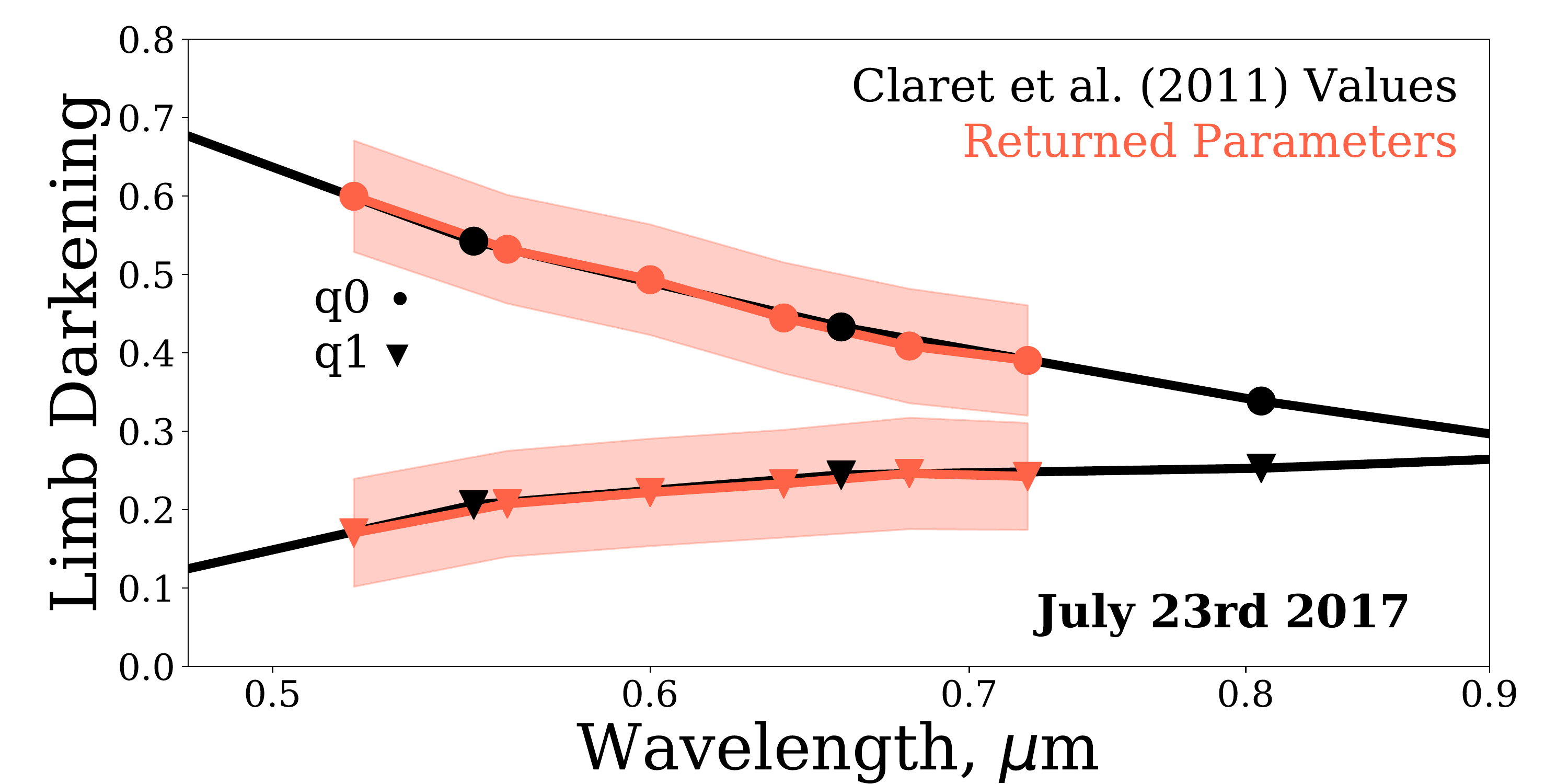}
    \plotone{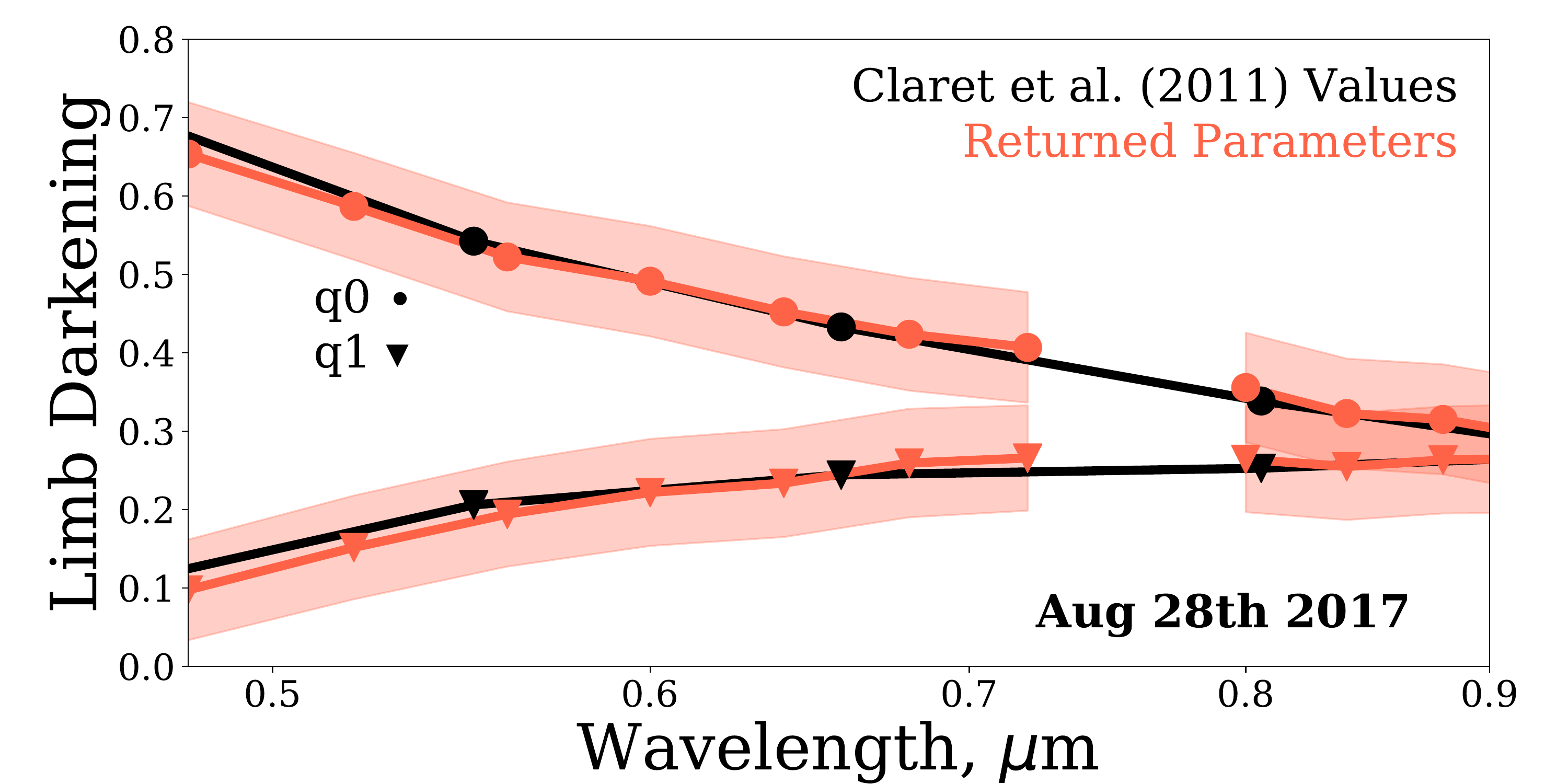}
    \caption{Limb darkening parameters from our MCMC fits. \textbf{Top:} Results from the night of July 23rd 2017. \textbf{Bottom:} Results from the night of August 28th 2017. For both, the theoretical values from \cite{Claret2011} are shown in black and our results in red. Points denoted by circles show the first quadratic coefficient, an upside down triangles show the second. Our 1-sigma uncertainty level is shown by the shaded region.} \label{limbdark}
\end{figure}

 Our initial quadratic limb darkening parameters are interpolated between the Johnson filter values from \cite{Claret2011}. While it is understood that these may not be completely accurate, our use of them as a starting value does not affect our results since we use a non-informative flat prior which allows our walkers to fully explore the parameter space without being biased to our initial guess. We confirm this by also starting white light chains at values of 0.0 for both limb-darkening parameters and the planet radius, which return the same results as those that begin at the \citeauthor{Claret2011} values. Because the unconstrained runs take longer to converge, we choose to use the starting guesses from \citeauthor{Claret2011} and a more informed initial guess for the planet radius. For this planet, we find that our returned limb darkening values agree well with those from \cite{Claret2011}. Figure \ref{limbdark} shows the theoretical quadratic limb darkening coefficient values and our fit values for both nights.
\section{Results} \label{results}
\subsection{Transmission Spectrum}
\begin{figure}
	\centering
    \epsscale{1.2}
    \plotone{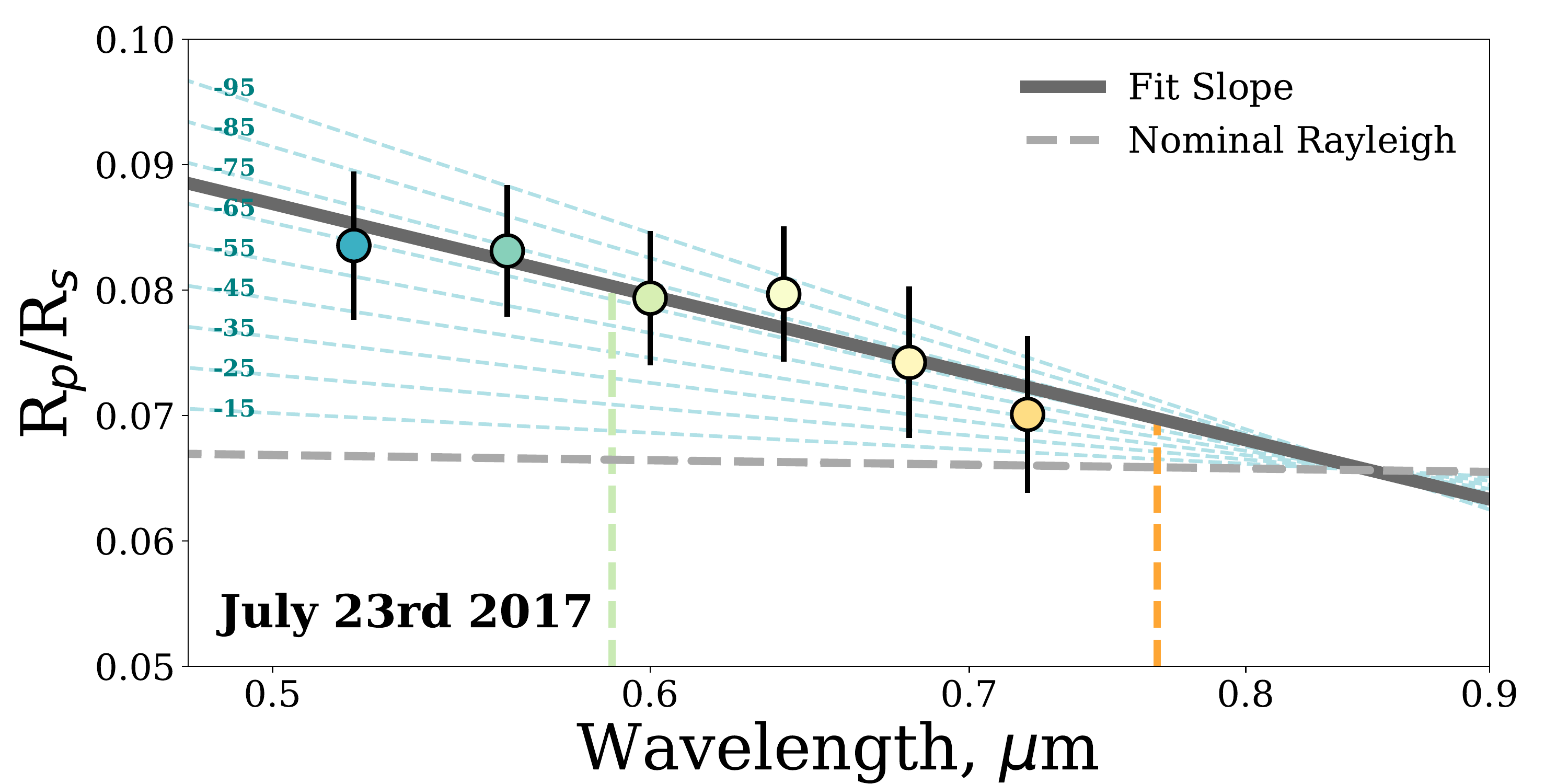}
    \plotone{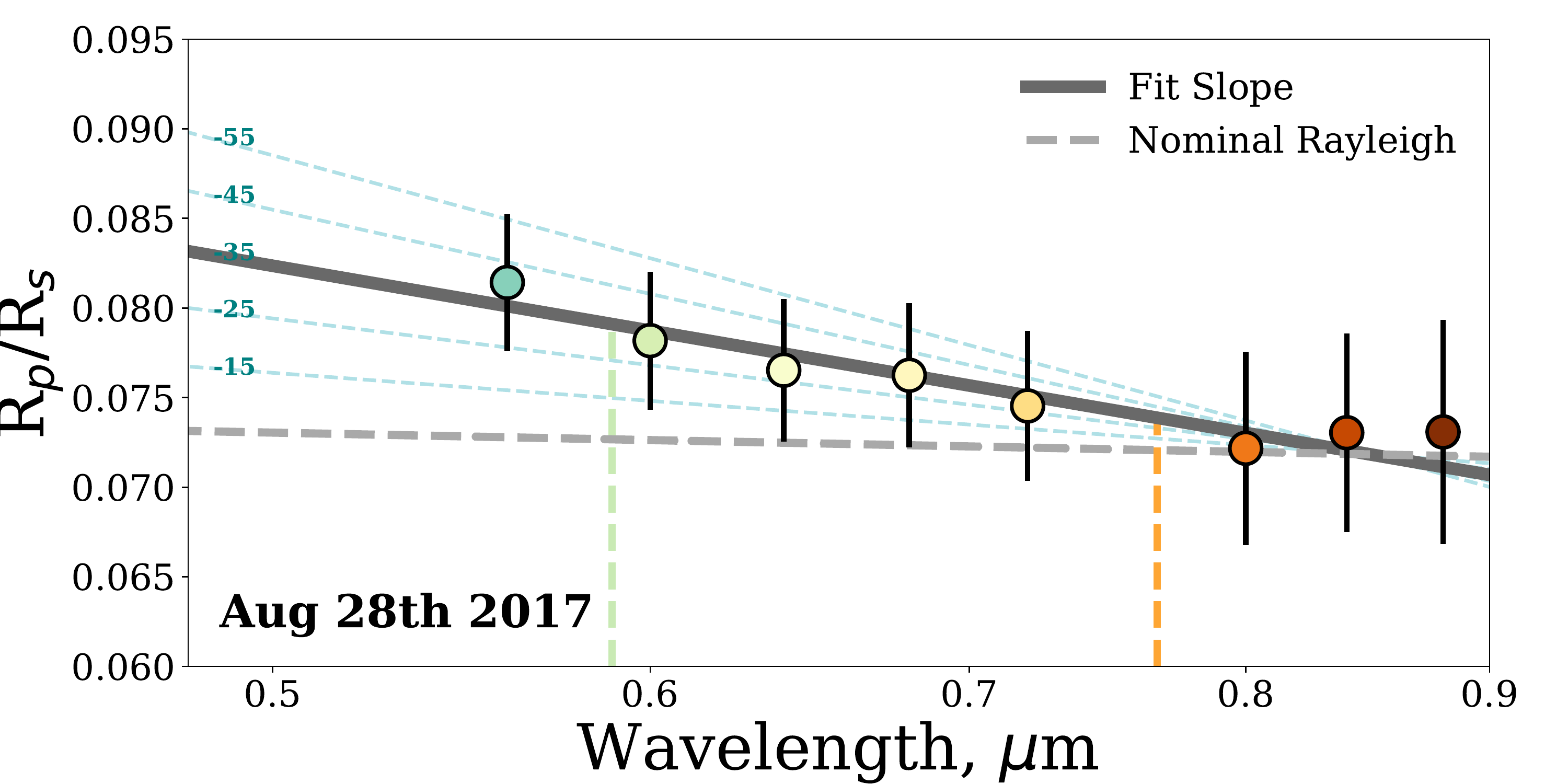}
    \caption{\textbf{Top:} Night one (July 23rd 2017) transmission spectrum. \textbf{Bottom:} Night two (August 11th 2017) transmission spectrum. The two nights are plotted separately due to differences in the measured scattering slope. On both panels, the vertical green dashed line at 5890\AA{ } is where we would expect sodium absorption, while the orange dashed line at 7665\AA{ } is where we would expect potassium absorption if they were present (though the features would not show up at the wide binning plotted here). The dashed gray line is the nominal Rayleigh slope, the solid gray line is the fit slope. Both panels include the same slopes for varying $\alpha$ parameters in blue for easy comparison of the two nights. The nominal Rayleigh slope has $\alpha=-$4, T$_{limb}$=T$_{eq}$ (1321 K), $\mu$=2.3, and g=g$_{planet}$ (464.51 cm/s$^2$.} \label{tspec}
\end{figure}

We fit our data from each night independently. We present results here for the 400\AA{ }binnings, which demonstrate the overall scattering slope. We find no definitive evidence of Sodium or Potassium absorption at any of our finer binnings. Figure \ref{tspec} shows our transmission spectra and scattering slopes. Table \ref{tableresults} lists our results for both nights.
\par Our data covers only optical wavelengths, which can only constrain combinations of atmospheric properties by itself. Due to the lack of absorption features, we report only measurements of the strength of the scattering slope. The optical slope we detect is typically a signature of scattering, given by 
\begin{equation}
	\frac{1}{R_S}\frac{dR_p}{d\ln\lambda}=\alpha\frac{1}{R_S}\frac{k_BT_{limb}}{\mu g}
    \label{rayleigh}
\end{equation}
first described in \cite{Lecavelier2008}, where R$_S$ is the stellar radius, R$_p$ is the planetary radius, $\alpha$ is the power of the wavelength dependence in the scattering cross section ($-$4 for Rayleigh Scattering), k$_B$ is the Boltzmann constant, T$_{limb}$ is the temperature at the limb of the planet, $\mu$ is the mean molecular weight of the atmosphere, and g is the planetary gravity. 1/R$_{S}$ is included on both sides of the equation because transmission spectra are typically plotted as the planet radius relative to the stellar radius. 
\par Our main unknowns in Equation \ref{rayleigh} are the atmospheric mean molecular weight ($\mu$) and the limb temperature ($T_{limb}$). Typical values for $\mu$ range from 2.2 for Jupiter to 2.5-2.7 for Neptune. If the atmosphere of the planet is clear, the gas should produce Rayleigh scattering and $\alpha$ will be equal to $-$4 as on Earth. However, if there exist particles in the atmosphere large enough to produce Mie scattering, $\alpha$ can take on a wider range of values. 
\par Our measured transmission slope on night 1, July 23rd 2017,  is $({1}/{R_S})({dR_p}/{d\ln\lambda})=-$0.092$\pm$0.025, $\sim$17.5 times stronger than nominal Rayleigh scattering, defined as the right hand side of Equation \ref{rayleigh} with $
\alpha=-$4, T$_{limb}$=T$_{eq}$, $\mu=$2.3, and g=464.51 cm/s$^2$ (calculated from the planetary parameters given in Table \ref{Hats8_params}. With these parameters, there is no reasonable combination of $\mu$ and limb temperature to explain the slope at $\alpha=-$4.0. At the equilibrium temperature of the planet and a mean molecular weight of 2.3, we can match the slope with an extremely strong wavelength dependence of $\lambda^{-70}$ for scattering. 
\par On night 2, the measured slope is $({1}/{R_S})({dR_p}/{d\ln\lambda})=-$0.046$\pm$0.025, approximately one-half of that from night 1. Prior to searching parameter space for a combination of $\mu$, T$_{limb}$, and $\alpha$ that matches this slope, we first discuss the effect unocculted star spots may have on the measured scattering slope during our night 1 observations (see section \ref{spots}) in order to explain the difference between the two slopes. After correcting night 1 to have a slope that matches night 2, we will explore options to explain the remaining strong scattering slope. 
\par An important consideration is how uncertainties on planetary properties may affect the nominal Rayleigh slope, which in turn affects our expectations. Taking our parameters and uncertainties from Table \ref{Hats8_params}, we note that R$_{planet}$ has the highest relative uncertainty. When estimating the Rayleigh slope, planet radius matters only for the planet gravity. This results in a factor of R$_{planet}^2$. If the true planet radius sits at the edge of the upper error bar, so that R$_{planet}$=0.996$\times$R$_{Jupiter}$, the nominal Rayleigh slope will increase by a factor of 1.3. This is not a significant change when compared to the factor of $\sim$18 we are looking to explain and so we do not explore this option further. 
\par In addition, HATS-8b is a low density, low gravity, exoplanet, with $\rho$=0.259 g/cm$^3$, 2.6$\times$ lower than Saturn. It is possible that the extended atmosphere requires a different treatment for atmospheric models. In particular, Exo-Transmit \citep{Kempton2017} does not account for variations in scale height with altitude, a potentially important factor for these low-density, inflated planets.
\par As noted above, we also search for the presence of sodium absorption (5890\AA) and potassium absorption (7665\AA) with our narrow bins. We find that the presence or lack of potassium absorption could not be well established due to the deep O$_2$ telluric line at $\sim$7600\AA{ }diluting any signal. We do not find strong evidence of sodium absorption. 
\subsection{Unocculted Star Spots} \label{spots}
Although we do not detect an occulted spot during either of the observed transit events, we cannot rule out that the host star has some amount of spot coverage. Particularly because unocculted spots can appear as a scattering slope in transmission spectrum, and our results suggest a stronger than expected slope with differences between the two nights. If the rotational period of the host star is approximately twice the time between observations, we would be viewing the planet against a different side of the star, and different spot coverage. 
\par Following the approach of \cite{MOPSS1}, we use the formulation of \cite{Louden2017} to investigate the influence unocculted spots may have on our results. We assume that night 1 is heavily influenced by spots, while night 2 is relatively uninfluenced by spots because of the stronger slope measured on night 1. The measured transit depth, $\delta_m(\lambda)$, is described as a function of the true transit depth, $\delta(\lambda)$, the spot coverage fraction, $\eta$, and the relative flux from a spot, $F_{\lambda}$(spot), compared to the stellar photosphere, $F_{\lambda}$(star).
\begin{equation}
	\label{deltaradius}
	\delta_m(\lambda)=\delta(\lambda)\frac{1}{1-\eta\left(1-\frac{F_{\lambda}(spot)}{F_{\lambda}(star)}\right)}
\end{equation}
The overall dimming of the star due to unocculted spots can be written as
\begin{multline}
	F_{\lambda}(\mathrm{star,corrected})=\eta f_{\lambda} F_{\lambda}(\mathrm{star})+(1-\eta)F_{\lambda}(\mathrm{star}) \\ =[1-\eta(1-f_{\lambda})]F_{\lambda}(\mathrm{star})
\end{multline}
with $f_{\lambda}$ the spot contrast given as $F_{\lambda}$(spot)/$F_{\lambda}$(star). When the star dims due to unocculted spots, the transit depth looks relatively larger compared to transits with no unocculted spots. Because the spot's black body spectra peaks at longer (redder) wavelengths, while the star peaks at relatively shorter (bluer) wavelengths, the spot contrast level is higher at short wavelengths, and lower at red wavelengths. This is therefore a wavelength dependent effect, with blue wavelengths more heavily affected than red, resulting in a star-induced slope being injected into the data. 
\par
\begin{figure}
	\centering
    \epsscale{1.2}
    \plotone{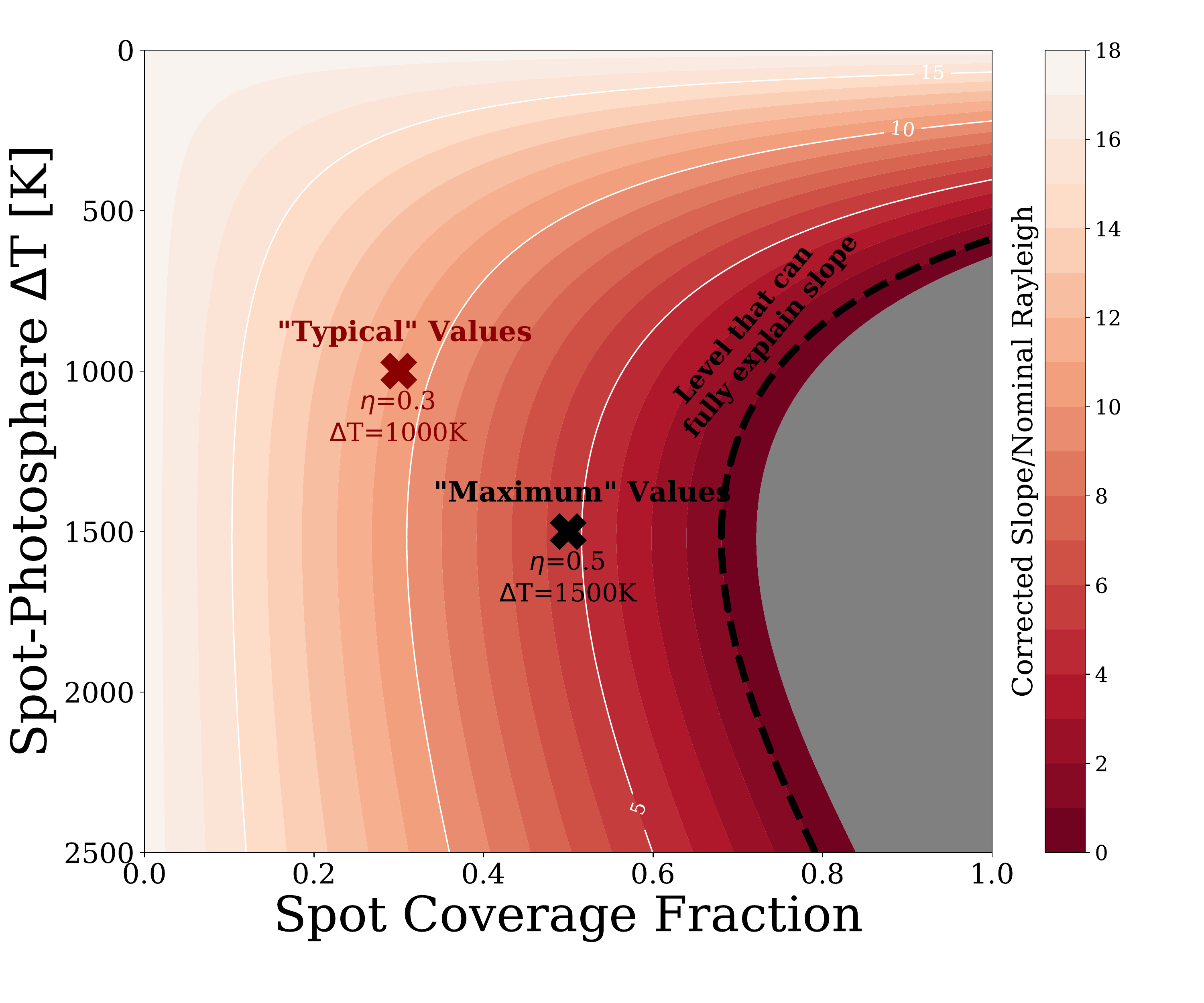}
    \plotone{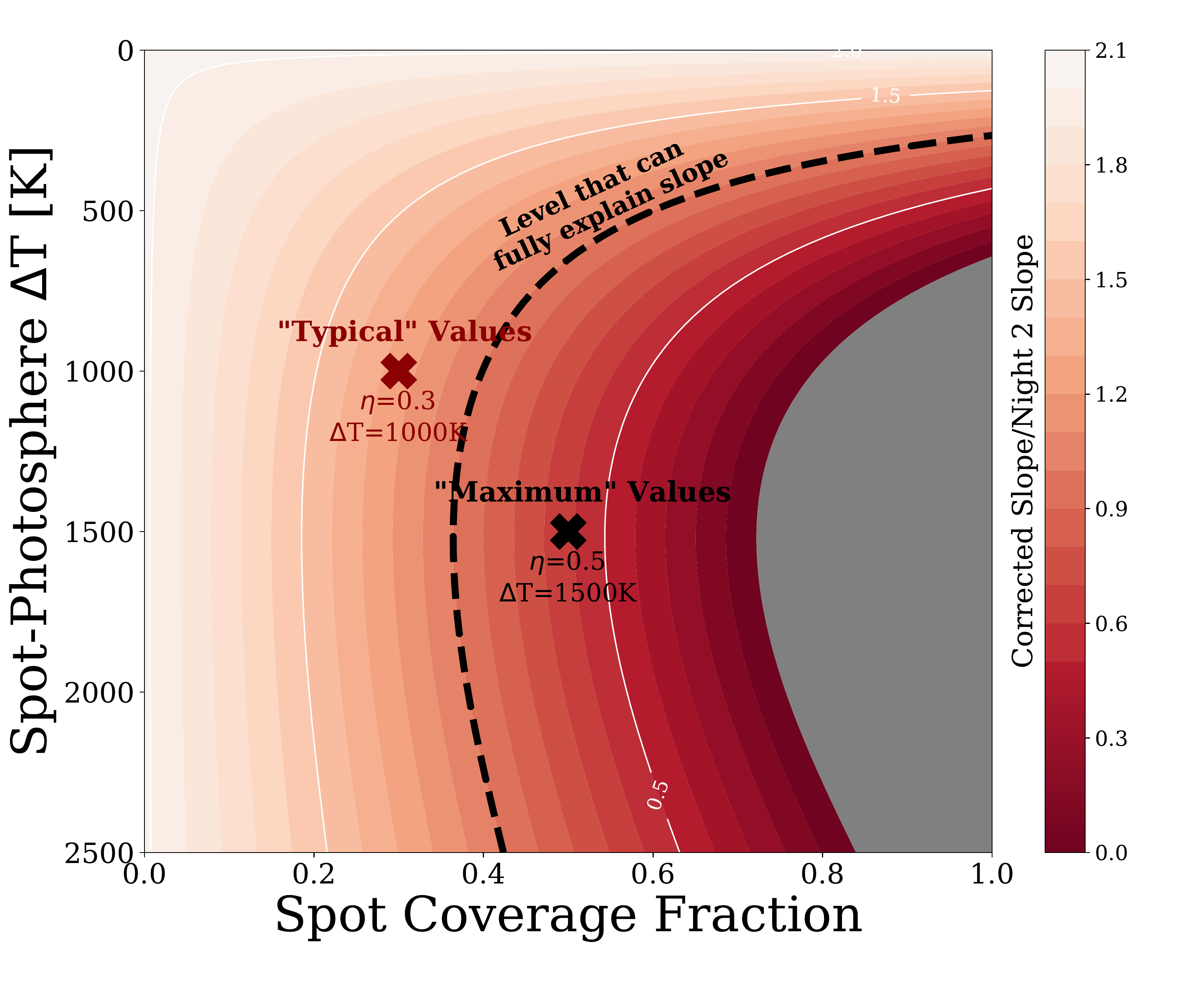}
    \caption{Equation \ref{deltaradius} applied to our observed slope for a variety of spot contrast and spot coverage fractions. \textbf{Top:} The corrected slope is divided by the nominal Rayleigh slope. \textbf{Bottom} The corrected slope is divided by the night 2 slope. In both panels, The black dashed line denotes the level that would correct our data back onto the nominal Rayleigh slope or night 2 slope, at a ratio of 1. The higher the value, the further the corrected slope is from the `goal' slope. We also mark with a red 'X' a representative ``typical'' set of values and a black 'X' the ``maximum'' spot coverage and $\Delta$T we consider in this work, these are the same values on both panels. The shaded gray regions result in a positive slope, and so are considered ``unallowed''. Regardless, it is highly unlikely that a star would have such a high spot coverage and $\Delta$T. We are unable to correct the slope back onto the nominal Rayleigh through unocculted spots alone as demonstrated in the top panel, but we can explain the difference between the two nights.} \label{unocc_plot}
\end{figure}

 Following equation \ref{deltaradius}, we calculate the required $\delta_m(\lambda)/\delta(\lambda)$ to match the observed night 1 slope to both the nominal Rayleigh slope and the observed slope on night 2. Figure \ref{unocc_plot} shows the corrected/expected slopes (where a ratio of 1 can fully explain the slope) at a variety of spot coverage fractions ($\eta$) and $\Delta$T=T$_{star}$-T$_{spot}$. 
\begin{splitdeluxetable}{cccccccBccccccc} 
\tabletypesize{\footnotesize}
\tablecolumns{14}
\tablecaption{HATS-8b: MCMC Fit Results \label{tableresults}}
\tablewidth{0.7\textwidth}
\tablehead{						&
    \multicolumn{7}{c}{July 23rd 2017\, {\large}}	&
    \multicolumn{5}{c}{August 28th 2017\, {\large}} \\
    \cline{2-7} \cline{9-14}
    \colhead{Bin Center}	&
    \colhead{R$_{p}$/R$_{star}$}	&
    \colhead{$\Delta$ R$_{p}$/R$_{star}$}	&
    \colhead{q0}	&
    \colhead{$\Delta$ q0}	&
    \colhead{q1}	&
    \colhead{$\Delta$ q1}	&
    \colhead{Bin Center}	&
    \colhead{R$_{p}$/R$_{star}$}	&
    \colhead{$\Delta$ R$_{p}$/R$_{star}$}	&
    \colhead{q0}	&
    \colhead{$\Delta$ q0}	&
    \colhead{q1}	&
    \colhead{$\Delta$ q1}	
    }
\startdata	
5200 $\AA $&	0.0836	&	0.0059	&	0.600	&	0.071	&	0.170	&	0.069	& 5200 $\AA $ &	---		&	---		&	---		&	---		&	---		&	---		\\
5600 $\AA $&	0.0831	&	0.0053	&	0.532	&	0.069	&	0.207	&	0.068	& 5200 $\AA $ &	0.0814	&	0.0038	&	0.522	&	0.069	&	0.194	&	0.067	\\
6000 $\AA $&	0.0794	&	0.0054	&	0.493	&	0.071	&	0.222	&	0.068	& 5200 $\AA $ &	0.0782	&	0.0039	&	0.491	&	0.070	&	0.222	&	0.068	\\
6400 $\AA $&	0.0797	&	0.0054	&	0.444	&	0.071	&	0.233	&	0.069	& 5200 $\AA $ &	0.0765	&	0.0040	&	0.452	&	0.071	&	0.234	&	0.069	\\
6800 $\AA $&	0.0742	&	0.0060	&	0.409	&	0.073	&	0.246	&	0.071	& 5200 $\AA $ &	0.0763	&	0.0040	&	0.424	&	0.072	&	0.259	&	0.069	\\
7200 $\AA $&	0.0701	&	0.0063	&	0.390	&	0.070	&	0.242	&	0.068	& 5200 $\AA $ &	0.0745	&	0.0042	&	0.407	&	0.071	&	0.266	&	0.067	\\
7600 $\AA $&	---		&	---		&	---		&	---		&	---		&	---		& 5200 $\AA $ &	---		&	---		&	---		&	---		&	---		&	---		\\
8000 $\AA $&	---		&	---		&	---		&	---		&	---		&	---		& 5200 $\AA $ &	0.0722	&	0.0054	&	0.356	&	0.070	&	0.264	&	0.068	\\
8400 $\AA $&	---		&	---		&	---		&	---		&	---		&	---		& 5200 $\AA $ &	0.0730	&	0.0056	&	0.323	&	0.070	&	0.255	&	0.068	\\
8800 $\AA $&	---		&	---		&	---		&	---		&	---		&	---		& 5200 $\AA $ &	0.0731	&	0.0063	&	0.315	&	0.070	&	0.263	&	0.068	\\
9200 $\AA $&	---		&	---		&	---		&	---		&	---		&	---		& 5200 $\AA $ &	0.0695	&	0.0080	&	0.295	&	0.071	&	0.265	&	0.069	\\
\enddata
\tablecomments{ Here q0 is the first quadratic limb darkening parameter and q1 is the second quadratic limb darkening parameter. Night 1 is reported prior to the unocculted star spot correction. Night 1 does not have values reported past Earth's 7600 O$_2$ absorption feature due to second order contamination. For both nights, the bin centered on this absorption feature is discarded.}
\end{splitdeluxetable}
\par The required spot coverage fraction to fully explain our night 1 slope vs. nominal Rayleigh slope at a $\Delta$T of 1500K is $\sim$75\%, much higher than one would expect. HATS-8 is a G star with an age of 5.1$\pm$1.7 Gyr \citep{Bayliss2015}, comparable to the Sun. On magnetically active low-mass stars, \cite{Jackson2013} suggest that the spot coverage could be as high at 40\% with T$_{spot}$/T$_{star}$=0.7 (similar to our value). Further, long term spot-coverage studies by \cite{Alekseev2018} find spot coverages up to 42\% for 13 active G and K stars. In agreement with these literature values for G stars, we can explain the difference between our two nights with a spot coverage fraction of 40\% and a $\Delta$T of 1500K. We note, however, that if HATS-8 had 40\% spot coverage on night 1, it is rather unlikely that the observed transit did not exhibit a spot crossing event unless the orbit and rotational axis of the star are very well aligned and the spots are strongly bound to latitudes away from the path of the transit. In addition, a 40\% spot coverage fraction on night 1 is even more unlikely because our two observations are only one month apart. Regardless, we make this correction in order to move forward in attempting to explain the large slope with a joint fit between the two nights and acknowledge the need for photometric monitoring of the star HATS-8.
\par
\begin{figure}
	\centering
    \epsscale{1.2}
    \plotone{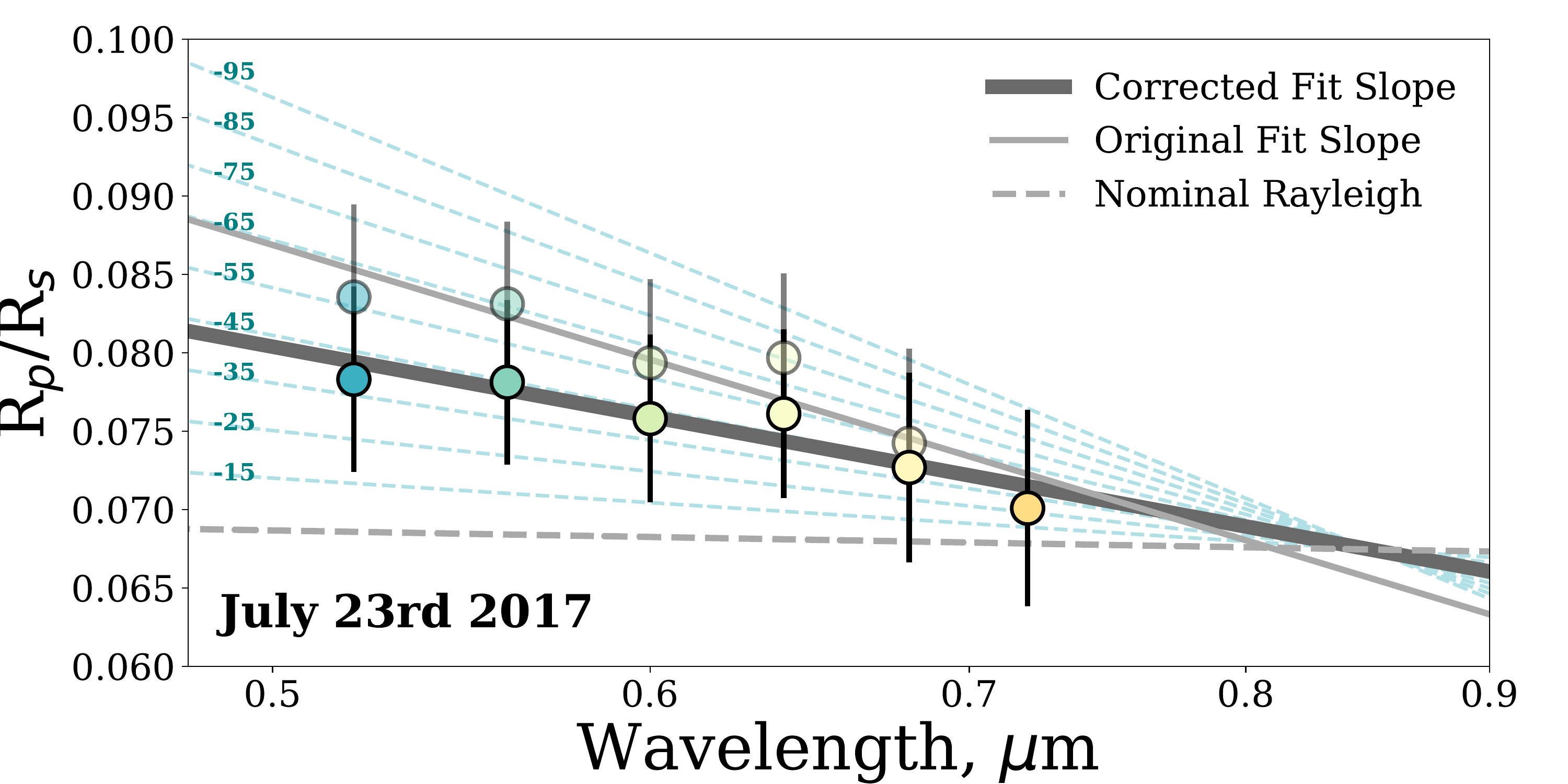}
    \plotone{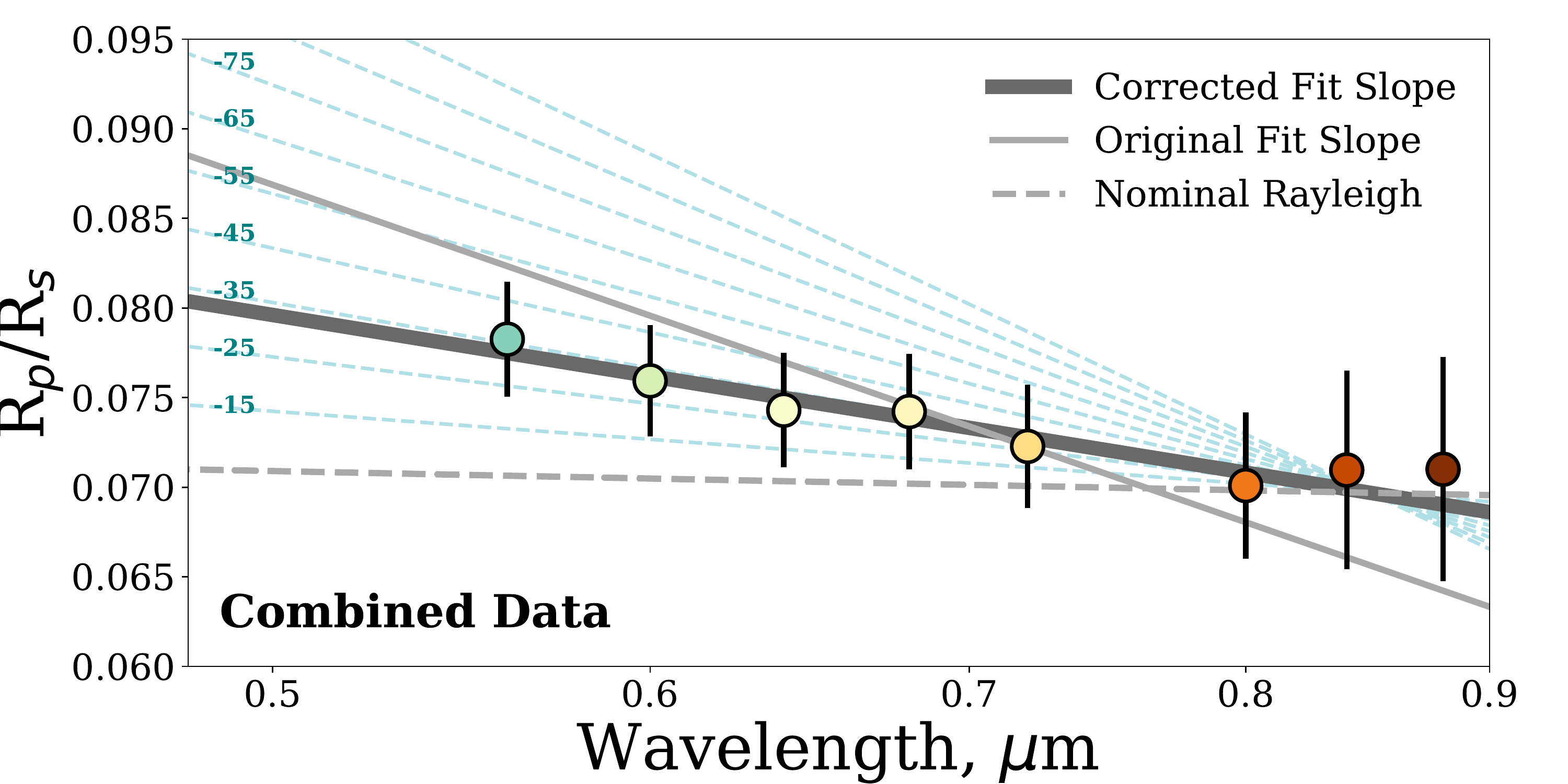}
    \caption{\textbf{Top}: Night 1 (July 23rd 2017) transmission spectrum after applying a correction for unocculted star spots. \textbf{Bottom}: Both Nights combined. We find we are able to explain approximately half of the extreme scattering slope due to unocculted star spots with a coverage fraction of 40\% and $\Delta T$=1500 K, which is on the edge of our expectations for these parameters. We include the same blue dotted lines as in Figure \ref{tspec} for easy comparison, each line is labeled by the $\alpha$ value required to produce the slope.} \label{corrected_tspec}.
\end{figure}

 In Figure \ref{corrected_tspec}, we show the resulting transmission spectra, normalized so that the last bin is unchanged, after applying the unocculted star spot correction for a spot coverage of 40\% and a $\Delta$T of 1500K. With the difference in slopes explained as unocculted star spots, we are able to combine the two nights of data. Figure \ref{corrected_tspec} shows the combined spectrum in the bottom panel. The remaining analysis in this paper is performed on this averaged transmission spectrum.
\par After applying the correction for unocculted star spots with $\eta$=0.4 and $\Delta T$=1500K K, our combined measured slope is $({1}/{R_S})({dR_p}/{d\ln\lambda})=-$0.0431$\pm$0.005, half that prior to applying this correction but still $\sim$8 times larger than nominal Rayleigh Scattering. Table \ref{tableresultscom} lists our results for the combined nights.
\par
\begin{figure}
	\centering
    \epsscale{1.2}
    \plotone{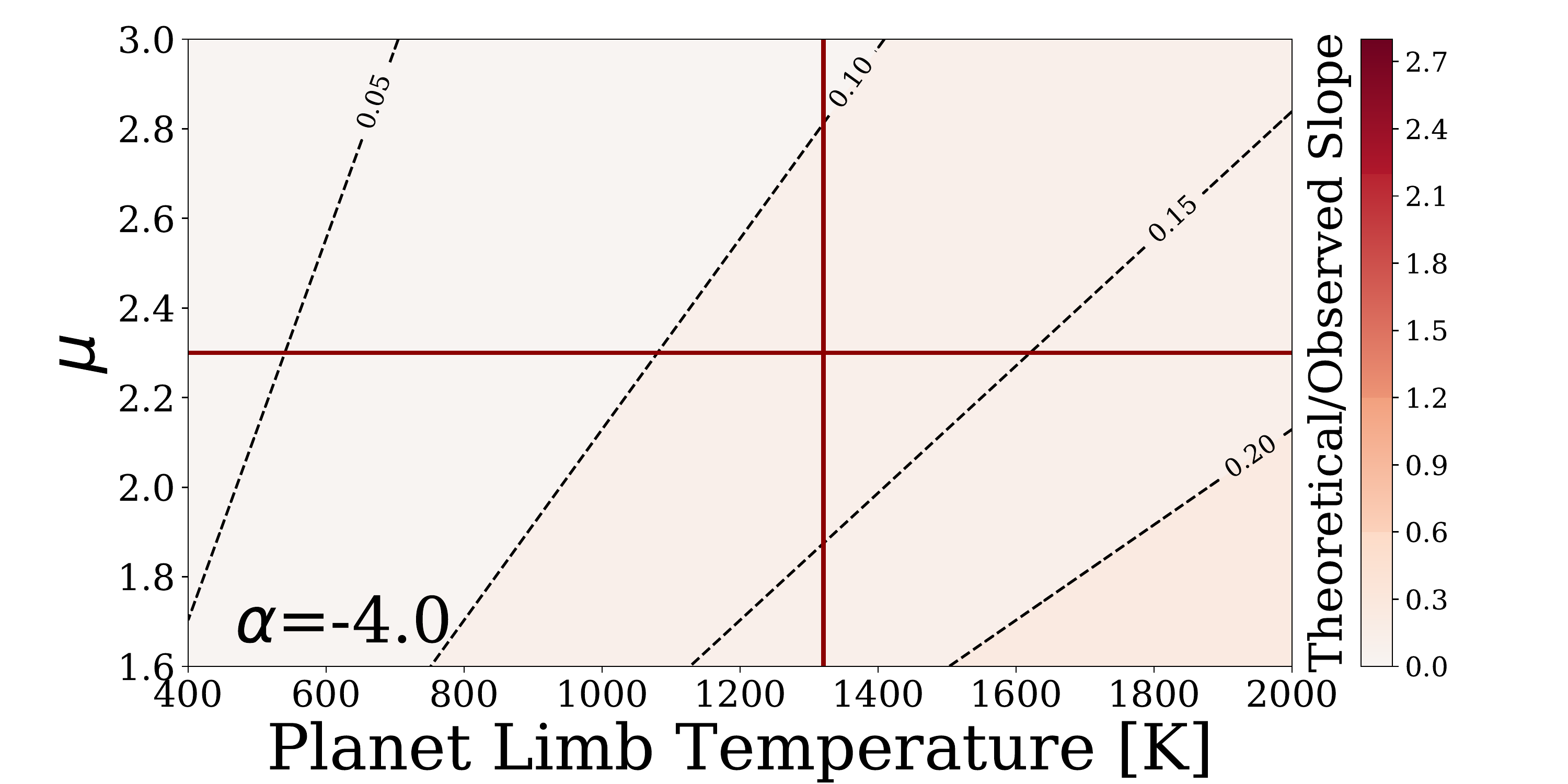}
    \plotone{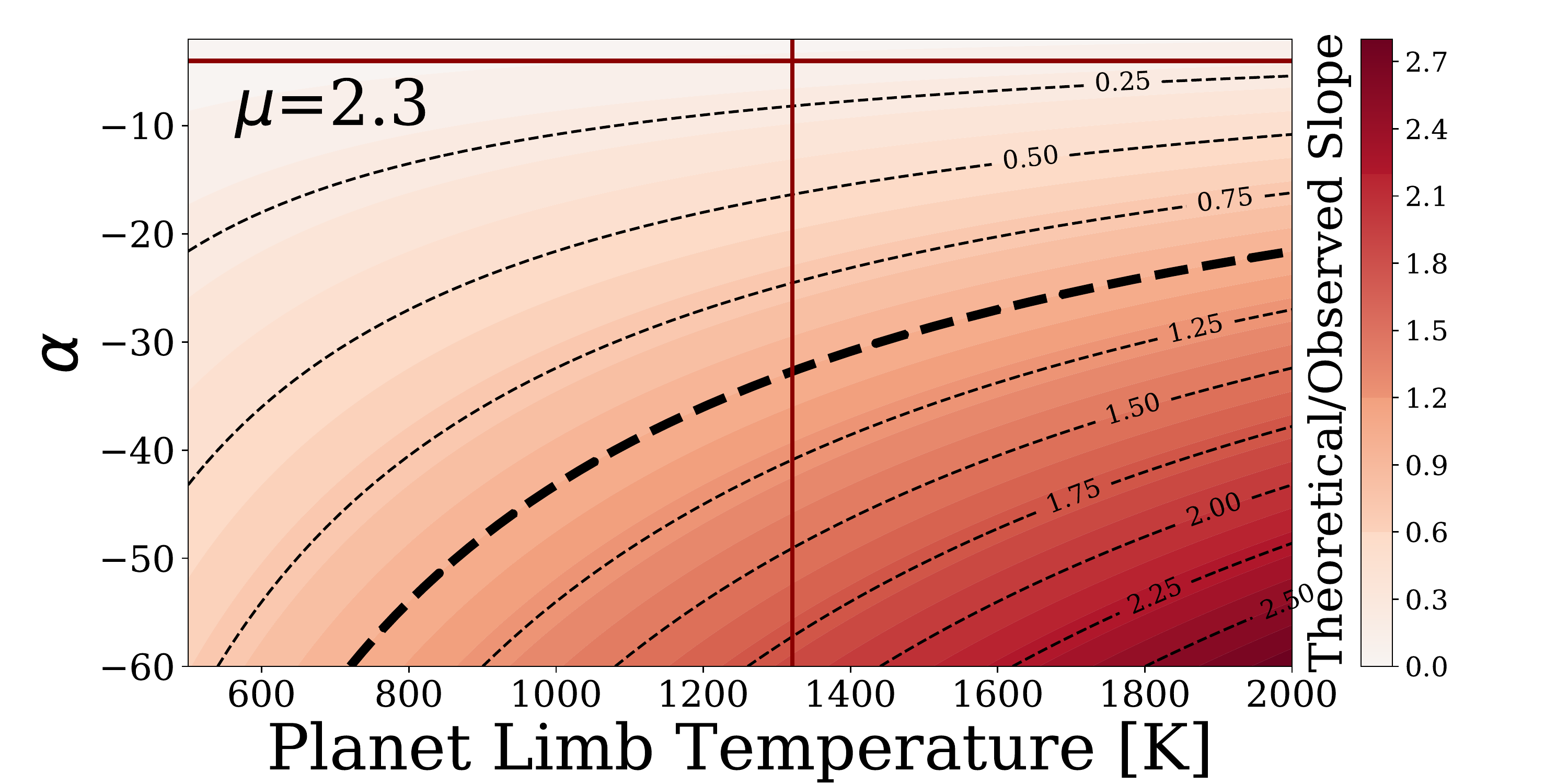}
    \plotone{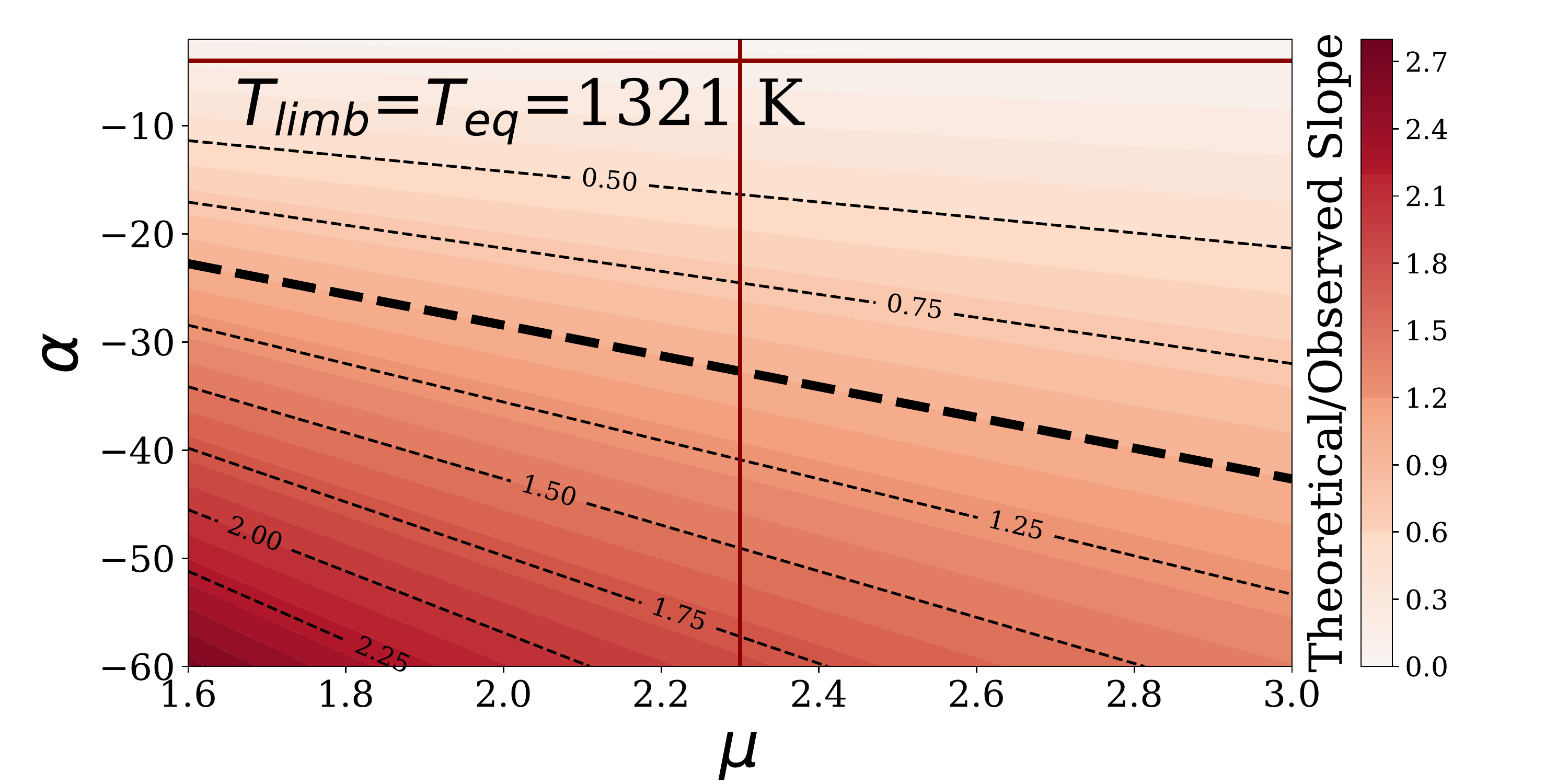}
    \caption{For all frames we show the theoretical Rayleigh slopes (see equation \ref{rayleigh}) divided by the measured slope after unocculted star spot correction, assuming a clear atmosphere. The dashed black line denotes the values that can explain the measured slope. Dark red lines represent the nominal values. \textbf{Top:} for a constant $\alpha$ and a range of planet limb temperatures and mean molecular weights. For a suitable range of these parameters, we find no combination that can explain our measured slope. \textbf{Middle:} for a constant mean molecular weight of 2.3 (a value typical of Jupiter) and a range of planet limb temperatures and $\alpha$. For T$_{limb}$=T$_{eq}\approx$1325 K, an $\alpha\approx$-35 would match the data. \textbf{Bottom:} for a constant limb temperature set to the equilibrium temperature and a range of $\alpha$ and $\mu$ values. The slope is not strongly sensitive to the mean molecular weight, but low values of $\mu$ allow for lower $\alpha$ values.} \label{slope_plots}.
\end{figure}

 With this slope, we explore the range of atmospheric parameters that can explain the observed slope in a clear atmosphere, after correcting for unocculted star spots. In Figure \ref{slope_plots} we show a range of possible parameters to satisfy the observed scattering slope in our data. Based on the unreasonable parameter vales shown in Figure \ref{slope_plots}, we require a strong wavelength dependence of approximately $\lambda^{-25}$ and a low mean molecular weight atmosphere ($\mu\lesssim$2.0) to explain our slope. This is inconsistent with a clear atmosphere, where $\alpha$ must equal $-$4.
\par As described below, we prefer an explanation that involves a portion of the slope due to clouds (see Section \ref{clouds}), and the remainder resulting from differences between our expected and realized values of mean molecular weight and the limb temperature.
\begin{deluxetable}{ccc}  
\tabletypesize{\small}
\tablecolumns{3}
\tablecaption{HATS-8b: MCMC Fit Results \label{tableresultscom}}
\tablewidth{0.9\textwidth}
\tablehead{
	\colhead{}						&
    \multicolumn{2}{c}{Combined}	\\
    \cline{2-3}
    \colhead{Bin Center [$\AA$]}	&
    \colhead{R$_{p}$/R$_{star}$}	&
    \colhead{$\Delta$ R$_{p}$/R$_{star}$}
    }
\startdata	
5200&	0.07568	&	0.00591	\\
5600& 0.07651	&	0.00310	\\
6000& 0.07383	&	0.00313	\\
6400& 0.07287	&	0.00320	\\
6800& 0.07184	&	0.00335	\\
7200& 0.07009	&	0.00348	\\
7600& ---		&	---		\\
8000& 0.06773	&	0.00539	\\
8400& 0.06860	&	0.00555	\\
8800& 0.06864	&	0.00626	\\
9200& 0.06501	&	0.00800	\\
\enddata
\tablecomments{The combined values account for both the unocculted star spot correction, and a relative shift between the two nights. Night 1 does not have values reported past Earth's 7600 O$_2$ absorption feature due to second order contamination. For both nights, the bin centered on this absorption feature is discarded.}
\end{deluxetable}
%
\subsection{Clouds and the Scattering Slope} \label{clouds}
Naturally, the next consideration is that the atmosphere of HATS-8b is not clear, and instead contains scattering particulates of some kind. The well known hot Jupiter, HD 189733b is a natural comparison here, due to its strong optical slope and similar equilibrium temperature (1200-1400K). A detailed analysis of the optical slope for HD 189733b is presented in \cite{Pont2013}. Of importance, however, is that \cite{Pont2013} state that the optical slope for HD 189733b can be explained by an $\alpha=-$4 Rayleigh scattering slope at a temperature of $\sim$1300K for particle sizes below 0.1$\mu$m. This is not the case for HATS-8b, as the slope is steeper than $\alpha=-$4 even after correcting for an extreme level of unocculted star spots. 
\par \cite{PandM2017} explore the effects of a variety of species on the wavelength dependence ($\alpha$) of the scattering cross section. In their work, \citeauthor{PandM2017} show that in some cases, the presence of cloud particles can result in a slope as strong as $\alpha=-$13. 
\par Of the species \citeauthor{PandM2017} discuss, they suggest that transmission spectra which show strong optical slopes are likely to contain sulphide clouds - Na$_2$S, MnS, or ZnS. These three species have condensation temperatures of 1176 K, 1139 K, and 700K, respectively at 10$^{-3}$ bar \citep{Wakeford2015}. It is probable that Na$_2$S and Mns could condense out in the nightside atmosphere of HATS-8b, which has an equilibrium temperature of 1321K for an albedo of 0. If a large amount of scattering particles are present, we would expect a higher albedo, and therefore lower temperature that allows the condensation of these species. Cloud formation along the limbs is an expected \citep{Parmentier2016,Roman2018} consequence of atmosphere dynamics.
\par Specifically, \citeauthor{PandM2017} find that MnS produces the steepest slopes at a modal particle size of 10$^{-2}$ $\mu$m for a cloud scale height (H$_c$) equal to the gas scale of k$_B$ T$_{eq}$/($\mu$ g). Under these conditions, one can attain a scattering slope with $\alpha=-$13 for a hot Jupiter of g = 24.79 m/s$^2$, $\mu$=2.3, and a solar abundance of MnS The slope models in \cite{PandM2017} depend on the cloud scale height (a smaller H$_c$ brings the slope closer to the standard Rayleigh slope), the modal particle size based on their assumed distribution (smaller particles cause steeper slopes), the reference pressure (only varies the absolute depth), the grain abundance, and the molecular abundance of the species. \citeauthor{PandM2017} state that the slopes are insensitive to the planet gravity, because the same value is assumed for both the the gas scale height and the cloud scale height. 
\par Additional condensates at these wavelengths and temperatures may include SiO$_2$, Al$_2$O$_3$, Fe$_2$O$_3$, Mg$_2$SiO$_4$, and MgSiO$_3$ with condensation temperatures of 1725 K, 1677 K, 1566 K, 1354K, and 1316 K, respectively at 10$^{-3}$ bar \citep{Wakeford2015}. Of these, \citeauthor{PandM2017} find that MgSiO$_{3}$ can result in the strongest scattering slope, with an $\alpha\sim-$5 for small grains (a modal particle size of 10$^{-2}\mu$m) if H$_c$=H. However, this is not much different from a nominal Rayleigh slope of $\alpha=-$4, and insufficient for the scenario we explore here. We note that this large slope may also be a result of a yet to be identified photochemical haze.
\par
\begin{figure}
	\centering    
    \epsscale{1.2}
    \plotone{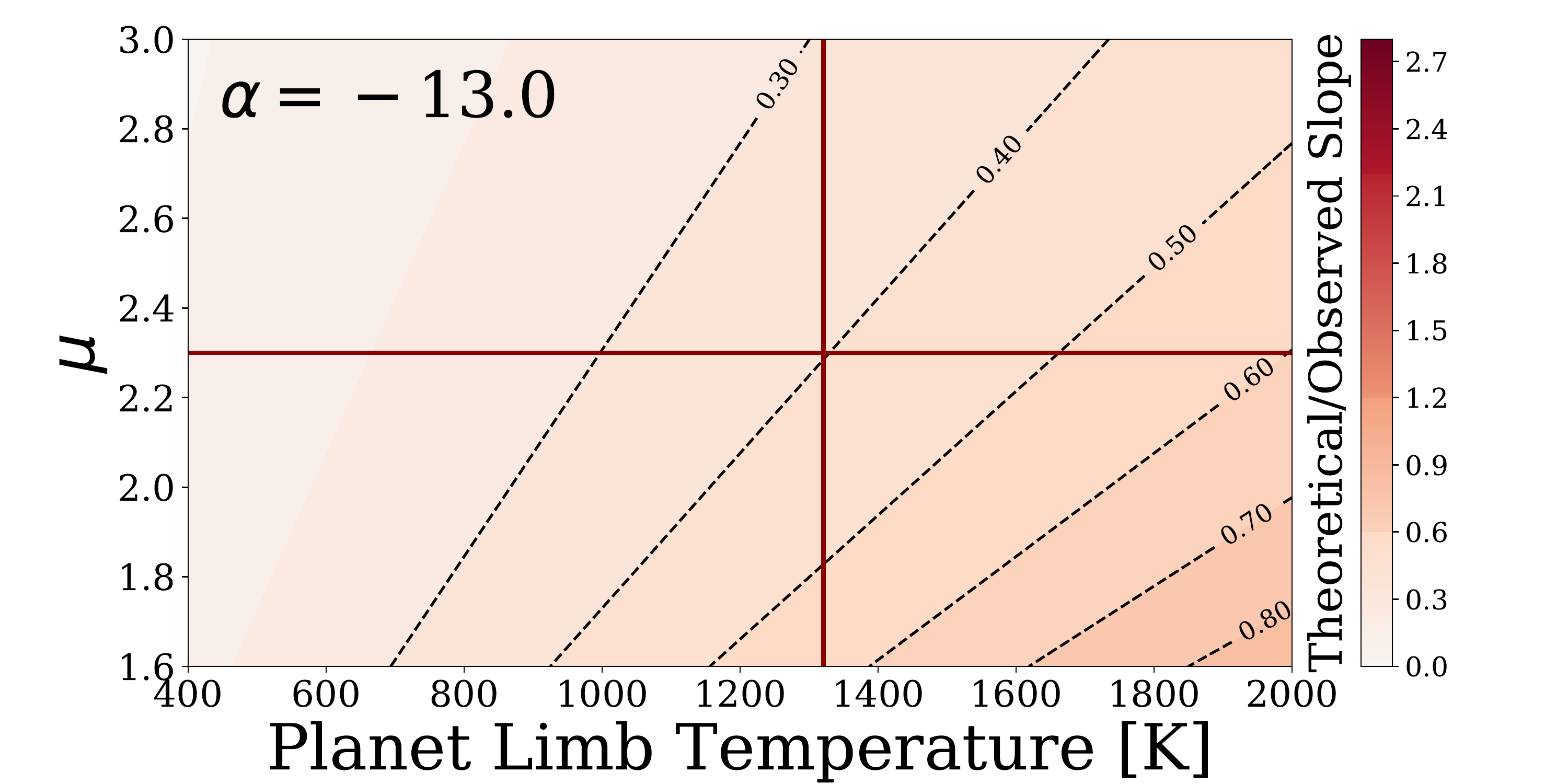}
    \caption{Here we show the necessary combinations of planetary limb temperature and mean molecular weight to match our observed data with an $\alpha$=-13, a value that may be attainable if MnS clouds are present with grain sizes of 10$^{-2}\mu$m. Dark red lines denote the nominal values.} \label{cloud_plot}.
\end{figure}

 In Figure \ref{cloud_plot} we show the values of limb temperature and mean molecular weight which, combined with $\alpha=-$13 for MnS clouds and small particle sizes, could explain our data. We plot the same levels as Figure \ref{slope_plots} for comparison. 
\par We find that we can explain at most 80\% of the observed slope through the inclusion of MnS clouds, but only for a warm, low mean molecular weight atmosphere. However, because a $\mu\textless$2.0 suggests a large fraction of the atmosphere is atomic rather than molecular, for reasonable values of $\mu\approx$2.0, we can at most explain $\sim$50\% of the slope at the equilibrium temperature of the planet.
\par Finally, In the limiting case that nothing is changing on the star itself, there must be a different explanation for the varying scattering slope we measure. Our two slopes differ by a factor of 3, which, following the methods of \cite{PandM2017}, suggests that the cloud scale height is up to 3 times larger on night 1. In order to get the strongest scattering slope ($\alpha=-$13) we assume H$_c$=H, which only gets us within a factor of 2 of matching our data. Invoking an increase in the vertical extent of the cloud to enhance the scattering would require vigorous vertical mixing in the atmosphere, but the change between the two observations would require that the mixing also be time variable, which has not been seen in hot Jupiter atmospheric simulations, to our knowledge. Therefore, if the particulates in the atmosphere of HATS-8b are composed of small MnS particles, it is not likely that we can explain the difference between the two nights as only an increase in cloud coverage on the terminator. 
\section{Conclusions}
\label{conclusions}
We observed the low density super-Neptune HATS-8b on two nights during August 2017 and September 2017. Both data sets show stronger-than-Rayleigh scattering, with significant differences between the two nights which can possibly be explained by unocculted star spots. After applying a correction to account for this possibility, we find the slope in the data from the combined two nights to be 27 times stronger than a nominal Rayleigh slope at the equilibrium temperature of the planet with a mean molecular weight of 2.3. We are unable to explain this slope with reasonable atmospheric parameters and gas scattering alone, so we explore several options to model the strong slope.
\begin{itemize}
\item We have explored the possible condensates that could contribute to the strong scattering slope detected. MnS clouds with particle sizes no greater than 10$^{-2}\mu$m can result in at most $\alpha=-$13. This can explain half of our observed slope for a reasonable temperature and mean molecular weight, or up to 80\% of the slope if the limb temperature is $\sim$2000K. No other condensates (based on work by \cite{PandM2017} and \cite{Wakeford2015}) produce an $\left|\alpha\right|\gtrsim$5.
\item Uncertainties on planet and stellar parameters can at most account for a factor of 1.3 in the slopes. We discard errors in system parameters as an explanation for the measured transmission slope.
\end{itemize}
\par While we cannot completely rule out an instrumental or reduction process systematic, we are confident that they cannot explain the entire slope seen in this work. IMACS has not been previously found to have such a strong blue-red systematic. These are the most recent transits with this instrument currently in the literature, but work by \cite{Espinoza2018} includes transits from April 2017 without strong blue-red biases. We are unaware of any significant instrument work done during the three months between these transit observations. Further, our previous work \citep{MOPSS1} does not show strong blue-red slopes and uses the same reduction pipeline as this work. MOPSS1 presents work that is in agreement with the literature, so we do not expect strong blue-red biases to be a result of our reduction process. 
\par Photometric monitoring of HATS-8 would determine the activity level of the star, allowing a more informed correction for unocculted star spots. Future transits of HATS-8b would confirm if this slope is a time-variable phenomenon, and simultaneous with photometric monitoring, would point to if it is primarily a result of stellar or planetary effects. We recommend future follow up from ground and space based telescopes to better explain the atmosphere of HATS-8b and compare results from various sources. 

\acknowledgements
We thank the staff at Las Campanas Observatory, without which we would be unable to carry out the observations presented in this work. Special thanks to Dave Osip who took the August 2018 data through a service observing program at Las Campanas, and Michael Roman who provided useful comments on condensates for this manuscript. This research has made use of the Exoplanet Orbit Database and the Exoplanet Data Explorer at exoplanets.org. \nocite{exo-org} 
\facility{Magellan:Baade}
\software{\\ Astropy \citep{astropy},
\\ batman \citep{Kreidberg2015},
\\ emcee \citep{FMackey2013},
\\ IPython \citep{ipython},
\\ Matplotlib \citep{matplotlib},
\\ NumPy \citep{numpy},
\\ SciPy \citep{scipy},
\\ SpectRes \citep{Carnall2017}}


\end{document}